\documentclass[a4paper,Times New Roman,oneside,chapterprefix=true,fontsize=12pt]{scrbook}

\usepackage{url}
\usepackage[utf8]{inputenc}   
\usepackage[T1]{fontenc}      
\usepackage[small]{caption2}
\usepackage[raggedright]{sidecap}     
\usepackage[usenames,dvipsnames]{color}
\usepackage{graphicx}
\usepackage{float}
\usepackage{url}
\usepackage{scrlayer-scrpage}

\newcommand{\beq}{\begin{equation}}
\newcommand{\eeq}{\end{equation}}
\newcommand {\bea} {\begin{eqnarray}}
\newcommand {\eea} {\end{eqnarray}}

\usepackage{tabularx} 
\newcolumntype{L}[1]{>{\raggedright\arraybackslash}p{#1}} 
\newcolumntype{C}[1]{>{\centering\arraybackslash}p{#1}} 
\newcolumntype{R}[1]{>{\raggedleft\arraybackslash}p{#1}} 

\usepackage{amsbsy}
\usepackage{setspace}
\usepackage{amsmath}
\usepackage{amsfonts}
\usepackage{pdfpages}
\usepackage{amssymb}
\usepackage[sort&compress,numbers]{natbib}
\graphicspath{ {figures/} }
\usepackage{array}

\setheadsepline{0.5pt}                    
\setkomafont{pagehead}{\normalfont}
\setkomafont{pagefoot}{\normalfont}
\setkomafont{pagenumber}{\normalfont}

\usepackage[Sonny2]{fncychap}

\usepackage{lmodern}

\usepackage[pdftex]{hyperref}
\hypersetup{
	linktoc=page,
        colorlinks=true,               
        citecolor=blue,                 
        linkcolor=blue,                
        urlcolor=blue,                 
        breaklinks=true,               
 }

\usepackage[nodayofweek]{datetime}

\newdateformat{mydate}{\twodigit{\THEDAY}{ }\monthname[\THEMONTH], \THEYEAR}

\begin{document}

\frontmatter
\begin{titlepage}
    \begin{center}
        \vspace*{0cm}
        
        \LARGE
        \textbf{\Huge{Aspects of Quantum Cosmology}}
        
        \vspace{0.7cm}
        \Large
        Doctor of Philosophy 
        
        \vspace{0.7cm}
        
        \textbf{SACHIN PANDEY}\\
        \text{\small Roll no. 13IP016}
        
        \vspace{0.2 cm}
        \textbf{Under the supervision of \\Prof. Narayan Banerjee}\\
                \vspace{1 cm}
        \textit{\small A thesis submitted in the partial fulfilment of the requirements for the award of the Degree of Doctor of Philosophy to}\\
 \vspace{0.7cm}
        \large{Department of Physical Sciences}\\ \vspace{0.2cm}
         \includegraphics[width=0.3\textwidth]{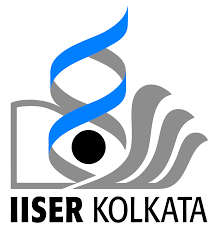}        
      
        \large{Indian Institute of Science Education and Research Kolkata}        
        
        \vspace{0.7 cm}
        December 2020
        
   \end{center}
\end{titlepage}

\begin{titlepage}
\newpage
\clearpage
\newpage
\mbox{~}
\clearpage
\newpage
\end{titlepage}
\thispagestyle{plain}
\null\vfil
{\huge\bfseries \centering Declaration \par\vspace{20pt}}
\hspace{10.5cm} Date : \mydate 30 December, 2020     \\

I, \textbf{Sachin Pandey} Registration No. \textbf{13IP016} dated 25 July 2013, a student of Department of Physical Sciences of the Integrated PhD Program of IISER Kolkata, hereby declare that this thesis is my own work and, to the best of my knowledge, it neither contains materials previously published or written  by  any  other  person, nor it has been submitted for  any  degree/diploma  or  any other academic award anywhere before. \\

I also declare that all copyrighted material incorporated into this thesis is in compliance  with the Indian Copyright (Amendment) Act, 2012 and that I have received written permission  from the copyright owners for my use of their work.  Chapter 2-6 of this thesis is based on work published in the peer-reviewed journals whose references are mentioned in respective chapters.

I hereby grant permission to IISER Kolkata to store the thesis in a database which can be accessed by others.

\vspace{2.0cm}\hspace{-0.5cm}
\vspace{0.05cm}
\textbf{Sachin Pandey} \\
Department of Physical Sciences \\
Indian Institute of Science Education and Research Kolkata\\
Mohanpur 741246, West Bengal, India
\begin{titlepage}
\newpage
\clearpage
\newpage
\mbox{~}
\clearpage
\newpage
\end{titlepage}
\thispagestyle{plain}
\null\vfil
{\huge\bfseries \centering Certificate \par\vspace{20pt}}
\hspace{10.5cm} Date : \mydate 30 December, 2020     \\

\vspace{0.0cm}
This is to certify that the thesis entitled \textbf{``Aspects of Quantum Cosmology"} submitted by Mr. \textbf{Sachin Pandey} Registration No. \textbf{13IP016} dated 25 July 2013, a student of Department of Physical Sciences of the Integrated PhD  Program of IISER  Kolkata, is based upon his own research work under my supervision. This is also to certify  that neither the thesis nor any part of it has been submitted for any degree/diploma or any  other academic award anywhere before. In my opinion, the thesis fulfils the requirement for  the award of the degree of Doctor of Philosophy.

\vspace{2.5cm}\hspace{-0.2cm}
\vspace{-0.0cm}
{\textbf{Prof. Narayan Banerjee}} \\
Professor \\
Department of Physical Sciences \\
Indian Institute of Science Education and Research Kolkata\\
Mohanpur 741246, West Bengal, India


\begin{titlepage}
\newpage
\clearpage
\newpage
\mbox{~}
\clearpage
\newpage
\end{titlepage}
\thispagestyle{plain}
\null\vfil
{\huge\bfseries \centering Acknowledgement \par\vspace{20pt}}
First and foremost, I would like to express my sincere gratitude to my supervisor Prof. Narayan Banerjee, for his constant support and advice throughout my doctorate studies. I also thank him for all the useful discussions and offering invaluable advices. He is always open to new ideas, and encouraged and helped me to shape my own interest and ideas. One simply could not wish for a better or friendlier supervisor. His words of encouragement sustained me through my PhD studies and  I am extremely grateful for that. I would never forget the stories he used to tell, especially the one where he mentioned, "Everything seems simple once it is done".

I would also like to thank my research progress committee members Dr. Golam Mortuza Hossain and Prof. Rajesh Kumble Nayak, for their valuable advice and support. Thanks to the Department of Physical Sciences, of IISER Kolkata for providing all the logistic support. I am thankful to Dr. Sridip Pal for insightful discussion. Thanks to CSIR(INDIA) for financial assistance through JRF-SRF(NET) fellowship during my PhD. I would like to thanks my teachers Dr. Rangeet Bhattacharyya from IISER Kolkata and Dr. Divya Haridas and Dr. Abhishek K. Singh and Prof. Patrick Das Gupta from the University of Delhi for their continuous support and valuable advice.

Completing this work would have been all the more difficult without the support and company of wonderful friends and seniors like Aabir Mukhopadhyay, Aakash Anand, Abhinash Kumar,  Ankit Kumar Singh, 
Dr. Anurag Banerjee, Avijit Chowdhury,  Debangana Mukhopadhayay, Deepak Kumar Jha, Dipanjan Chakraborty, Kaustav Gangopadhyay, Mayank Shreshtha, Nirbhay Kumar Bhadani,  Pawan Kumar, Purba Mukherjee, Tanima Duary, Dr. Prateek Verma, Dr. Preethi Thomas, Shibendu Gupta Choudhury, Shivanand Mandraha, Dr. Shubham Chandel, Ravi Kumar, Dr. Sayak Ray, Dr. Sajal Mukherjee,  Shashwat Kumar Singh, Soumik Mitra, Soumyajit Seth, Srijita Sinha, Sushant Kumar Sinha, Dr. Swati Sen; they made my years at IISER Kolkata an unforgettable experience. I am thankful to my college and school friends Aastha, Abhishek, Akhileshwar, Amit, Anuj, Anurag, Arif,  Atul,  Gaurav, Jewel, Meenal, Vipul for their continuous support and motivation.  \\I would also like to thank Ek Pehal kids with whom I have spent the memorable evenings and have learned from them while trying to teach them basic math and science.
 
Finally, I thank my parents and my sister for their unconditional love and support.
At the end of this fairly long testimonial, I once again thank everyone from whom I have learned something or the other...
\vskip 2cm
\emph{\\Sachin Pandey}\\
\emph{IISER Kolkata,\\ \mydate 30 December, 2020}
\clearpage 
\thispagestyle{plain}
\null\vfil
{\huge\bfseries \centering List of Publications \par\vspace{20pt}}
Content of this thesis is based on the following publications:
\begin{itemize}
\item Sachin Pandey, \textit{Anisotropic n-dimensional quantum cosmological model with fluid}, Eur. Phys. J. C \textbf{79}, 487 (2019). 

\item Sachin Pandey, Sridip Pal and Narayan Banerjee, \textit{Equivalence of Einstein and Jordan frames in quantized anisotropic cosmological models}, Annals Phys. \textbf{393}, 93-107 (2018).

\item Sachin Pandey and Narayan Banerjee, \textit{Equivalence of Jordan and Einstein frames at the quantum level }, Eur. Phys. J. Plus. \textbf{132}, 107 (2017). 

\item Sachin Pandey and Narayan Banerjee, \textit{Unitary evolution for anisotropic quantum cosmologies: models with variable spatial curvature}, Phys. Scripta \textbf{91}, 115001 (2016). 

\item  Sachin Pandey and Narayan Banerjee, \textit{Unitarity in quantum cosmology: symmetries protected and violated },  \href{http://arxiv.org/abs/arXiv:1911.11839}{\underline{arXiv:1911.11839}} (Submitted for the publication).\\  
\end{itemize}

\begin{titlepage}
\newpage
\clearpage
\newpage
\mbox{~}
\clearpage
\newpage
\end{titlepage}

\thispagestyle{plain}
\null\vfil
{\huge\bfseries \centering Abstract \par\vspace{20pt}}
\vspace{-0.5cm}
 In this thesis, we try to resolve the alleged problem of non-unitarity for various anisotropic cosmological models.  Using Wheeler-DeWitt formulation, we quantized the anisotropic models with variable spatial curvature, namely Bianchi II and Bianchi VI.  We used the dynamical variable related to the fluid as a time parameter following Schutz's formalism. We showed that Hamiltonian of respective models admits self-adjoint extension, thus unitary evolution. We further extended the unitary evolution for higher dimensional anisotropic cosmological models. We obtained finite normed and time independent wave packet showing the unitary evolution of the models. Expectation values of scale factors showed that the quantized model avoids the problem of singularity. We also showed that unitarity of the model preserves the Noether symmetry but loses the scale invariance.
 
 Using the Wheeler-DeWitt quantization scheme, we showed the equivalence of  Jordan and Einstein frames at the quantum level. We quantized the Brans-Dicke theory in both the frames for the flat FRW model using dynamical variable related to the scalar field as a time parameter. Obtained expressions for wave packet matched exactly in both the frames indicating the equivalence of frames. We also showed that equivalence holds true for various anisotropic quantum cosmological models, i.e., Bianchi I, V, X, LRS Bianchi-I and Kantowski-Sachs models.

\begin{titlepage}
\newpage
\clearpage
\newpage
\mbox{~}
\clearpage
\newpage
\end{titlepage}

\tableofcontents

\addcontentsline{toc}{chapter}{List of figures} 
\listoffigures
 
\listoftables
\addcontentsline{toc}{chapter}{List of tables}  
\begin{titlepage}
\newpage
\clearpage
\newpage
\mbox{~}
\clearpage
\newpage
\end{titlepage}
\mainmatter 

\chapter{Introduction}\label{ch:intro1}
A quantum theory of gravity should prescribe the quantum description of the universe. However, in the absence of a more generally accepted quantum theory of gravity, quantum cosmology is a moderately ambitious attempt to apply quantum principles in gravitational systems. Another strong motivation of quantum cosmology is the expectation that it might resolve the beginning of the universe from a singular state which plagues the classical description of the universe. This issue is discussed extensively in work by Hawking and Ellis\cite{hp},  Penrose\cite{penrose}, and Belinsky-Khalatnikov-Lifshitz  \cite{bkl}. Dynamics near the initial singularity is not well known even today.
                                   
 Wheeler-DeWitt equation\cite{dewitt1967quantum,wheeler1987superspace} provides a procedure of quantization which uses the Hamiltonian formulation in $3+1$ decomposition of gravitational configuration. It gives the evolution of the universe in infinite dimensional space of all possible three-geometries known as wheeler superspace. The problem to deal with infinite degrees of freedom is simplified in quantum cosmology, where the imposition of symmetries reduces infinite dimensional space to finite one, which is called minisuperspace.  
Quantum cosmology has its own sets of conceptual problems. The first one is the identification of time as our well known time parameter is a part of dynamics as a coordinate in relativistic theory.  There have been several approaches to address this issue like the use of internal coordinates, i.e., volume element or scalar field or external factors like matters field leading to some time parameter for the evolution of the universe. The method developed by  Lapchinskii and Rubakov \cite{Lapchinskii} uses the monotonic evolution of fluid density as a time parameter based on Schutz's formalism, which is discussed in detail in this thesis. 

Another major problem of quantum cosmology is a non-unitary evolution of the anisotropic universe \cite{alvarenga2002quantum,alvarenga2003, Majumder2013}, which questions the Wheeler-DeWitt formulation. The possibility to resolve this problem has been initiated for a few anisotropic models with the help of the self-adjoint extension of the Hamiltonian for the anisotropic models like Bianchi-I, V, IX \cite{spal1,spal2,spal3,spal4}. The first part of this thesis aims to deal with alleged non-unitarity in quantum cosmological models, especially anisotropic models with variable spatial curvature.

The latter part of the thesis deals with the equivalence of Jordan and Einstein frames in non-minimally coupled scalar field theories, particularly in Brans-Dicke theories  at the quantum level.

In this thesis following notations have been used:
Greek indices range over 0,1,2,3,4... and Latin indices range over 1,2,3....
A comma as in $A,_k$  denotes derivative of A with respect to $x^k$.
The signature convention of  + - - -  for the metric $g_{\mu \nu}$ is adopted.
Curvature tensors are used in following forms:
\bea
R^\tau_{\mu \nu \sigma}=\Gamma^\tau_{\mu\sigma,\nu}-\Gamma^\tau_{\mu\nu,\sigma}+\Gamma^\gamma_{\mu\sigma}\Gamma^\tau_{\gamma\nu}-\Gamma^\gamma_{\mu\nu}\Gamma^\tau_{\gamma\sigma},\\
R_{\mu \nu}=R^\tau_{\mu\tau\nu},\\
R=g^{\mu\nu}R_{\mu\nu},\\
\eea
where
\beq \Gamma^\tau_{\mu\nu}=\frac{g^{\tau\gamma}}{2}(g_{\nu\gamma,\mu}+g_{\mu\gamma,\nu}-g_{\mu\nu,\gamma}),\eeq
\beq
 g^{\mu\gamma}g_{\gamma\nu}=\delta^\mu_\nu.
\eeq
The choice of units, unless otherwise specified, will be $ \hbar =c =16\pi G =1$, where G is gravitational constant, $\hbar=\frac{h}{2\pi}(Planck's constant)$ and c is the speed of light in vacuum.
\section{Canonical formulation of Gravity}\label{ch:intro11}

The line element for the space-time defined on four dimensional manifold $\mathcal{M}$ characterized with metric tensor $g_{\mu\nu}(x^{\gamma})$ assigned on coordinate $x^{\gamma}$ can be given as
\beq 
ds^2=g_{\mu\nu}dx^\mu dx^\nu.
\eeq
Einstein-Hilbert action which contains the gravitational contribution and matter contribution is 
\beq \label{1action}
A=\int_{\mathcal{M}}d^4x(\sqrt{-g}R+\mathcal{L}_m).
\eeq
We need to introduce the Hamiltonian for the gravitational field for a canonical quantization given by DeWitt \cite{dewitt1967quantum}. This requires the $3+1$ split of the metric in which spatial components of metric are dynamical degrees of freedom. In geometrical language,  this can be considered as a decomposition of the four-dimensional manifold $\mathcal{M}$ as  $M \rightarrow R \times \Sigma $, where $\Sigma $ denotes a three-dimensional hypersurface. Arnowitt, Deser, Misner(ADM)\cite{adm} describe the usual 3 + 1 split through the decomposition of the space-time metric in terms of a lapse function $N$ , a shift vector $N_i$ and an induced spatial metric $h_{ij}$ . One can begin by constructing the hypersurfaces $\Sigma_t$ , which is parameterized by some global time-like variable $t$.

In the 3+1 decomposition metric tensor is given as,
\beq
g_{\mu\nu}=\begin{pmatrix}
N^2-N^kN_k & -N_j \\
-N_j & -h_{ij}.
\end{pmatrix}
\eeq
The extrinsic curvature of a hypersurface $\Sigma_t$ is defined as
\beq
K_{ij}=\frac{1}{2N}(N_{i|j}+N_{j|i}-h_{ij,0}),
\eeq
where vertical bar denotes the covariant derivative with respect to three dimensional metric $h_{ij}$.
\textit{Gauss-Codazzi relation} relates the  four dimensional Ricci scalar $R$ to three dimensional one $^3R$ with the help of the extrinsic curvature $K_{ij}$,
\beq 
\sqrt{-g} R=N\sqrt{h}(^3R+K_{ij}K^{ij}-K^2)-2(\sqrt{h}K)_{,0}+2(\sqrt{h}KN^j\sqrt{h}-h^{ij}N_{,i})_{,j}, \label{1lag}
\eeq
where $h=det(h_{ij})$, $K^{ij}=h^{ik}h^{jl}K_{kl}$ and $K=h^{ij}K_{ij}$.\\
The last two terms of Eq.(\ref{1lag}) represent a derivative on the boundary in the action and do not contribute to dynamics and thus may be dropped.
Now the gravitational part of the action(\ref{1action}) can be written in $3+1$ formalism as
\beq
A_g(N,N^i,h_{ij})=\int_{M}dt dx^3\mathcal{L}_g=\int_M N\sqrt{h}(^3R+K_{ij}K^{ij}-K^2)dtdx^3 \label{1action3d}.
\eeq
We can write conjugate momenta to the variables ($N, N^i,h_{ij} $) from  the Lagrangian density in the following way,
\bea
\Pi=\frac{\delta\mathcal{L}}{\delta\dot{N}}=0, \label{1cm1}\\
\Pi_i=\frac{\delta\mathcal{L}}{\delta\dot{N}^i}=0, \label{1cm2}\\
\Pi^{ij}=\frac{\delta\mathcal{L}}{\delta\dot{h}_{ij}}=-\sqrt{h}(K^{ij}-h^{ij}K).\label{1cm3}
\eea
Equations (\ref{1cm1}) and (\ref{1cm2}) are known as \textit{primary constraints}. They express the fact that the Lagrangian does not depend on the \textit{velocities} $\dot{N}$ and $\dot{N}_{i}$.
\\Hamiltonian can be written in terms of $N, N^i, h_{ij}$ and their conjugate momenta using the \textit{Legendre transformation} as
\beq
H=\int d^3x (\Pi\dot{N}+\Pi_i\dot{N}^i+\Pi^{ij}\dot{h}_{ij}-L),
\eeq
\beq
H=\int d^3x (\Pi\dot{N}+\Pi_i\dot{N}^i+N^i\mathcal{H}_i+N\mathcal{H}),
\eeq 
where
\bea
\mathcal{H}=\mathcal{G}_{ijkl}\Pi^{ij}\Pi^{kl}-\sqrt{h} (^3R), \label{1shn}\\
\mathcal{H}_i=-2h_{ik}\Pi^{kj}_{|j}, \label{1smm}\\
\mathcal{G}_{ijkl}=\frac{1}{2\sqrt{h}}(h_{ik}h_{jl}+h_{jk}h_{il}-h_{ij}h_{kl}). \label{1smc}
\eea
$\mathcal{H},\mathcal{H}_i$ and $\mathcal{G}_{ijkl}$ are called as \textit{super Hamiltonian, super momentum} and \textit{super-metric or DeWitt metric} respectively \cite{dewitt1967quantum}.
\\ In terms of these variables, the action for gravity part(\ref{1action3d}) becomes
\beq
A_g=\int dt dx^3 (\Pi\dot{N}+\Pi_i\dot{N}^i-N^i\mathcal{H}_i-N\mathcal{H}).
\eeq 
If we vary the above action with respect to $\Pi^{ij}$, we get back the relation given in equation(\ref{1cm3}). Variation of action with respect to lapse function N gives the Hamiltonian constraint
\beq
\mathcal{H}=0,
\eeq
while variation of the action with respect to shift vector $N^i$ gives the supermomentum constraint
\beq
\mathcal{H}_i=0
\eeq
These constraints can also be obtained as 
\bea
\dot{\Pi}=\frac{\partial(N\mathcal{H})}{\partial N}=\mathcal{H}=0, \label{1sc1}\\
\dot{\Pi_i}=\frac{\partial(N^i\mathcal{H}_i)}{\partial N^i}=\mathcal{H}_i=0 \label{1sc2}.
\eea 
They are also known as \textit{secondary constraints} in \textit{Dirac} terminology.
\section{Quantization Scheme: Wheeler-DeWitt Equation}\label{ch:intro2}
The \textit{super Hamiltonian} as given in equation(\ref{1sc1}) provides the evolution of the system. The \textit{supermomentum constraint} defines the configurational space of the canonical gravity that it is the infinite dimensional space of all the possible geometries\cite{Wiltshire:1995vk}. 
\subsection{Superspace}\label{ch:intro2a}
For an infinite-dimensional space $x=\{ x^i\}$, which specifies the point on the hypersurface $\Sigma$, there are finite number of degrees of freedom at each point, one considers the space of all Riemannian 3-metric configuration on the spatial hypersurfaces,
\beq
Riem(\Sigma)=\{h_{ij}(x), \hspace{1.5cm} x \in \Sigma\}.
\eeq
Since we are interested in the geometry and configuration related to each other by a diffeomorphism, we identify the \textit{superspace}\cite{wheeler1987superspace} as
\beq \frac{Riem(\Sigma)}{Diff_0(\Sigma)},\nonumber \eeq
where zero subscript denotes the diffeomorphism connected to the identity.
Then, the DeWitt metric (\ref{1smc})  can be written as
\beq
\mathcal{G}_{AB}(x)=\mathcal{G}_{(ij)(kl)}(x),
\eeq
where the indices $A, B \in (h_{11},h_{12},h_{13},h_{22},h_{23},h_{33})$, run over all the independent components of intrinsic metric $h_{ij}$. Signature of DeWitt metric at each point x is always (- + + + + +), regardless of the signature of the space-time metric $g_{\mu\nu}$. We can also extend the range of indices to include the matter fields by defining appropriate $\mathcal{G}_{\phi\phi}(x)$.

Classical Poisson brackets for the configurational variables  can be written as
\beq
\{h_{\alpha\beta}(x,t),h_{\gamma\delta}(x',t)\}=0,\eeq 
\beq
\{\Pi^{\alpha\beta}(x,t),\Pi^{\gamma\delta}(x',t)\}=0,\eeq\beq
\{h_{\alpha\beta}(x,t),\Pi^{\gamma\delta}(x',t)\}=0,
\eeq 
and the corresponding quantum Poisson brackets can be given as
\beq
\{\hat{h}_{\alpha\beta}(x,t),\hat{h}_{\gamma\delta}(x',t)\}=0,\eeq \beq
\{\hat{\Pi}^{\alpha\beta}(x,t),\hat{\Pi}^{\gamma\delta}(x',t)\}=0,\eeq \beq
\{\hat{h}_{\alpha\beta}(x,t),\hat{\Pi}^{\gamma\delta}(x',t)\}=0.
\eeq 
The quantum operators corresponding to the classical variables can be written as
\bea
N\rightarrow \hat{N} ,\hspace{0.9cm} N_i\rightarrow \hat{N}_i ,\hspace{0.9cm} h_{ij}\rightarrow \hat{h}_{ij} , \\
\Pi\rightarrow \hat{\Pi} =-i\frac{\delta}{\delta N},\hspace{0.9cm} \Pi^i\rightarrow \hat{\Pi}^i =-i\frac{\delta}{\delta N_i}, \hspace{0.9cm} \Pi^{ij}\rightarrow \hat{\Pi}^{ij} =-i\frac{\delta}{\delta h_{ij}}.
\eea
Let us consider the wave function of the universe to be a functional state $\Psi(N,N_i,h_{ij})$, which is annihilated by the quantum version of the primary constraint(\ref{1cm1},\ref{1cm2}) as
\beq
 \hat{\Pi}\Psi =-i\frac{\delta \Psi}{\delta N} =0,\hspace{1cm} \hat{\Pi}^i\Psi =-i\frac{\delta \Psi}{\delta N_i}=0.
\eeq
These relations establish the fact that the wave function of the universe is independent of $(N,N_i)$ and become a functional of three metric $h_{ij}$ only. Similarly, using supermomentum constraint, we get
\beq
 \hat{\mathcal{H}}^i=2i \left(\frac{\delta \Psi}{\delta h_{ij}}\right)_{|j}=0.
\eeq
The above relation implies the fact that the wave function of the universe does not depend on a particular metric used to represent the geometry. Instead, it is defined on the whole class of the three geometries.  This can be expressed as
\beq
\Psi=\Psi\{h_{ij}\}.
\eeq
This indicates that wave function exists in the same configurational space, i.e., superspace, as discussed in the beginning.
Finally, the  quantum counterpart of the super Hamiltonian constraint read as
\beq
\hat{\mathcal{H}}\Psi =\mathcal{G}_{ijkl}\frac{\delta^2 \Psi}{\delta h_{ij} \delta h_{kl}}-\sqrt{h}\hspace{0.2cm}^3R \Psi =0. \label{1wdw}
\eeq
Equation (\ref{1wdw}) is known as \textit{Wheeler-DeWitt}(WDW) equation\cite{dewitt1967quantum,wheeler1987superspace}. It is the second-order differential equation at each point $x \in \Sigma$ on the superspace. We can see the ambiguity of factor ordering in this WDW equation, but there exist natural choices of ordering for which derivative terms become a Laplacian in the supermetric.
\subsection{Minisuperspace}\label{intro2b}
In section \ref{ch:intro2a}, the canonical quantization method considers the configurational space of the infinite dimensional space of all possible three-geometries. The presence of an infinite number of degrees of freedom makes the problem intractable with the techniques that have been developed so far. It is necessary to truncate the infinite degrees of freedom to finite numbers from a practical perspective. One can obtain some particular minisuperspace model through the imposition of symmetries in superspace \cite{Kuchar:1989tj,Sinha:1991nc,Mazzitelli:1992ng,Ishikawa:1993bv}. Considering homogeneous metrics is an easy choice to achieve this.
Let us consider a spatially homogeneous system characterized by zero shift vector $N^i=0$, only time dependent lapse function $N(t)$ given as
\beq
ds^2=N^2(t)dt^2-h_{ij}dx^idx^j. \label{1mini}
\eeq 
In this minisuperspace, the three metric $h_{ij}$ depends on the finite number of coordinates $q^A$, unlike the superspace case where we have the infinite dimensional degree of freedom.

The Einstein-Hilbert action for this minisuperspace can be written as 
\beq
\mathcal{A}=\int dt[\frac{1}{2N}\mathcal{G}_{AB}(q)\dot{q}^A\dot{q}^B-NU(q)],
\eeq
where $\mathcal{G}_{AB}$ is the minisupermetric, the reduced version of the entire supermetric $\mathcal{G}^{ijkl}$ given as
\beq
\mathcal{G}_{AB}dq^Adq^B=\int d^3x \frac{1}{2}\mathcal{G}^{ijkl}\delta h_{ij}\delta h_{kl},
\eeq
 and $U(q)$ is the potential term given as
 \beq
 U=\int d^3x \sqrt{h} (-^3R).
 \eeq 
Since our configurational space is finite dimensional, quantization is simplified as we are dealing with the quantum mechanics of the constrained system. 

Now canonical momenta and Hamiltonian can be obtained as
\bea
\Pi_A=\frac{\partial L}{\partial \dot{q}^A}=\frac{\mathcal{G}_{AB}\dot{q}^B}{N}, \\
H=\Pi_A\dot{q}^A-L=N(\frac{1}{2}\mathcal{G}^{AB}\Pi_A\Pi_B+U(q)).
\eea
We can obtain the super Hamiltonian constraint by varying the action with respect to the lapse function as
\beq
\mathcal{H}=\frac{1}{N}H=\frac{1}{2}\mathcal{G}^{AB}\Pi_A\Pi_B+U(q)=0. \label{1mha}
\eeq
With natural choices of ordering \cite{Hawking:1985bk}, canonical quantization of equation(\ref{1mha}) leads to  the Wheeler-DeWitt equation as
\beq
\hat{\mathcal{H}}\Psi=[-\frac{1}{2}\nabla^2+U(q)]\Psi=0,
\eeq
where the symbol $\nabla$ denotes the covariant derivative constructed from minisupermetric and the Laplacian $\nabla^2$ for the minisupermetric is given as
\beq \label{1mswdw}
\nabla^2=\frac{1}{\sqrt{-\mathcal{G}}}\partial_A[\sqrt{-\mathcal{G}}\mathcal{G}^{AB}\partial_B],
\eeq
where $\mathcal{G}=det(G_{AB})$.
\section{The Problem of Time }\label{ch:intro3}
The quantum Hamiltonian constraint $\mathcal{H}$ (\ref {1sc1}) is that the operator $\hat{\mathcal{H}}$  annihilates the physical states. The well known Schrödinger equation for our  Hamiltonian constraint equation can be written as 
\beq
\hat{\mathcal{H}}\Psi=i\frac{\partial \Psi}{\partial t}=0.
\eeq
This equation denotes the time independence of the wave function $\Psi$, which apparently discard the quantum evolution of the system. Such frozen formalism\cite{DERUELLE1989253} seems to suggest that the quantum theory of gravity does not evolve with the time.  This issue is known as the \textit{problem of time} \cite{anderson2012problem}.

Difference between the intrinsic nature of the two theory, general relativity, and quantum mechanics give rise to this issue. In quantum mechanics, the time is used for the evolution of the system, and events occur on it as time is an externally fixed scalar parameter. 

The absence of a suitable time parameter creates countless problems in quantum theory. Any quantum theory is unable to explain the concepts of probability and measurement in the absence of proper time description unless one redefines these concepts.
In order to avoid the problem of time, we must have a first order momenta term in Hamiltonian. One can identify a time coordinate $q_A$ at the classical level. Then one can write scalar constraint with the help of  the conjugate momenta $p_A$ to the time variable as
\beq
p_A+H_A=0,
\eeq 
where $H_A$ is the physical reduced Hamiltonian which can evolve with respect to the time variable $q_A$ and subsequently Schrödinger like equation can be written as
\beq
\hat{H}_A\Psi=i\frac{d\Psi}{dq_A}.
\eeq
The selection of time variable $q_A$ can be made in two ways. First, one can select $q_A$ from gravitational configurational space. But with this selection, one can only quantize a part of gravitational space. The quantization of the reduced ADM Hamiltonian using this scheme is shown by Arnowitt, Deser, and Misner \cite{adm}. 

The second possibility can be a selection of the time coordinate $q_A$ from the external matter field and evolve the gravity part of the Hamiltonian with respect to  time, which is picked up from the evolution of the matter distribution. One such procedure is the use of the fluid so that the monotonic evolution of the fluid density can be identified as a time parameter. Lapchinskii and Rubakov \cite{Lapchinskii} uses this procedure with the help of the \textit{Schutz's formalism}, and it is discussed in detail in the following section.
\section{Schutz's Formalism}\label{ch:intro4}
In hydrodynamics, a perfect fluid can be described with the help of velocity potentials \cite{sw}. Schutz \cite{schutz1970perfect} generalized it as a nonlinear relativistic field theory for five coupled scalar fields, whose Lagrangian density is simply the pressure of the fluid. A similar generalization was also independently suggested by Schmid \cite{schmid}.  In this section, we discuss the conversion of perfect-fluid hydrodynamics into the Hamiltonian form, as given by Schutz \cite{schutz1971hamiltonian}.

Action for fluid part can be written as 
\beq
A_f=\int dt d^3x \sqrt{-g}\mathcal{P}=\int dt d^3x N\sqrt{h}\mathcal{P}, \label{1fact}
\eeq
 where $\mathcal{P}$ is pressure of fluid related to density $\rho$ by an equation of state ($\mathcal{P}=\alpha\rho$) .
 Now let us express the four velocity $U_\nu$ as
\beq 
 U_\nu=\frac{1}{\mu}\left(\epsilon_{,\nu}+\zeta \beta_{,\nu}+\theta S_{,\nu} \right),\label{1fv}
\eeq
where  $S$ is specific entropy, $\zeta$ and $\beta$ are potentials connected with rotation, and $\epsilon$ and $\theta$ are potentials with no clear physical meaning and $\mu$ is specific enthalpy given as
\beq
\mu=\frac{\mathcal{P}+\rho}{\rho_0}=1+\mathcal{u}+\frac{\mathcal{P}}{\rho}. \label{1enth}\eeq
Here $\rho_0$ is rest mass density and $\mathcal{u}$ is specific internal energy. 
The four velocity is normalized as 
\beq
U_{\nu}U^{\nu}=1. \label{1nfv} 
\eeq
From the laws of thermodynamics, one can write 
\beq
\tau(\rho_0, \mathcal{u})dS=d\mathcal{u}+\mathcal{P}d(1/\rho_0)=(1+\mathcal{u})d(ln(1+\mathcal{u})-\alpha \hspace{0.2cm}ln\rho_0).
\eeq
From above relation, expression for entropy $S$ can be given as
\beq
S=ln(1+\mathcal{u})-\alpha\hspace{0.2cm} ln\rho_0. \label{1entro}
\eeq
Using equations (\ref{1enth}) and (\ref{1entro}), the fluid pressure can be given as
\beq
\mathcal{P}=\alpha \frac{\mu^{1+1/\alpha}}{(1+\alpha)^{1+1/\alpha}}e^{-\frac{S}{\alpha}}. \label{1fp}
\eeq
In co-moving system, the four-velocity $U_\nu=(N,0,0,0)$, then equations (\ref{1fv}) and (\ref{1nfv}) yield as
\beq
\mu=\frac{1}{N}(\dot{\epsilon}+\zeta\dot{\beta}+\theta\dot{S}). \label{1sfv}
\eeq
Using equations (\ref{1fp}) and (\ref{1sfv}), action for the fluid part(\ref{1fact}) can be given as 
\beq
\label{1factf}
{\mathcal A}_f =\int d^3x dt {\mathcal L}_{f}\\ = \int dtd^3x \left[N^{-\frac{1}{\alpha}}\sqrt{h}\frac{\alpha}{\left(1+\alpha\right)^{1+\frac{1}{\alpha}}}\left(\dot{\epsilon}+\zeta\dot{\beta}+\theta\dot{S}\right)^{1+\frac{1}{\alpha}}e^{-\frac{S}{\alpha}}\right].
\eeq
If $q^a$ stand for the five potential fields $\epsilon, \zeta, \beta,\theta$ and $S$, then their conjugate momenta can be given as
\beq
p_a=\frac{\partial \mathcal{L}_f}{\partial \dot{q}^a}.
\eeq
This gives
\bea
p_{\epsilon}&=&N^{-\frac{1}{\alpha}}\sqrt{h}\frac{1}{\left(1+\alpha\right)^{\frac{1}{\alpha}}}\left(\dot{\epsilon}+\zeta\dot{\beta}+\theta\dot{S}\right)^{\frac{1}{\alpha}}e^{-\frac{S}{\alpha}}, \\
p_{\zeta}&=&p_{\theta}=0,\\
p_{\beta}&=&\zeta p_{\epsilon},\\
p_{S}&=&\theta p_{\epsilon}.
\eea
 Potential fields $\zeta $ and $\beta$ are related to the vorticity of the system. They do not contribute to any space-time without rotation. Hence they and their respective conjugate momenta can be dropped. This shows that we only have one independent momentum.

The Lagrangian density for the fluid part read as 
\beq \label{1lf}
\mathcal{L}_f=N^{-\frac{1}{\alpha}}\sqrt{h}\frac{\alpha}{\left(1+\alpha\right)^{1+\frac{1}{\alpha}}}\left(\dot{\epsilon}+\theta\dot{S}\right)^{1+\frac{1}{\alpha}}e^{-\frac{S}{\alpha}}.
\eeq
Since $p_\epsilon$ is only independent momentum, we can introduce the following canonical transformation
\bea \label{1can}
T&=&-p_{S}e^{-S}p_{\epsilon}^{-\alpha -1},\\
p_{T}&=&p_{\epsilon}^{\alpha+1}e^{S},\\
\epsilon^{\prime}&=&\epsilon+\left(\alpha+1\right)\frac{p_{S}}{p_{\epsilon}},\\
p_{\epsilon}^{\prime}&=&p_{\epsilon}.
\eea 
The corresponding Hamiltonian for the fluid part can be obtained from (\ref{1lf}) using above canonical transformed variable  as
\beq
H_f = N \frac{1}{\sqrt{h^\alpha}}p_T.
\label{1hf}
\eeq
The Poisson bracket $\{T,p_T\}$ corresponds to the quantum commutator in a canonical quantization, $p_T=-i\frac{\partial}{\partial T}$ as
\beq
[T,p_T]=i.
\eeq
The classical Poisson bracket $\{T,H_f\}$ gives 
\beq
\{T,H_f\}=\frac{1}{N}\frac{dT}{dt}= \frac{1}{\sqrt{h^\alpha}}.\label{1hpos}
\eeq
Since $\frac{1}{\sqrt{h^\alpha}}$ is a positive quantity, parameter $T$ can have the same orientation as the cosmic time everywhere in the system. Hence, parameter $T$ can be chosen as a time parameter, as discussed in the previous section. 
By varying the action with respect to $N$, one can get the super Hamiltonian $\mathcal{H}=\frac{\sqrt{h^\alpha}}{N}(H_g+H_f)=0$ which reads as 
\beq
\mathcal{H}_g+p_T=0.
\eeq
Now the Wheeler-DeWitt equation ($\hat{\mathcal{H}}\Psi=0$) reads as 
\beq
\hat{\mathcal{H}}_g\Psi=i\frac{\partial \Psi}{\partial T}. \label{1scwdw}
\eeq
Equation (\ref{1scwdw}) looks like the Schrödinger equation. Also T, which we identify as time, is indeed a scalar parameter and not a coordinate, unlike the cosmic time t. Thus the \textit{problem of time}, as discussed in section \ref{ch:intro3} is now resolved. Schutz's  formalism not just solves the problem of time, but also help to quantize the full part of gravitational space as mentioned in previous section \ref{ch:intro3}. As long as we can have 3+1 decomposition of metric of homogeneous system, this formalism should give us time parameter having same orientation as cosmic time(\ref{1hpos}).
\section{Homogeneous Cosmological Models}
\label{ch:intro5}
We notice regularity in the distribution of matter and radiation in our universe if we look at a large scale. An observable universe at a large scale suggests that our universe is homogeneous and isotropic everywhere. This phenomenon supported by observation is also known as \textit{Cosmological principle}. Isotropy means that there are no particular directions in the universe, i.e., all direction looks similar. Homogeneity means that there exist no specific or preferred places in our universe.

Roberston-Walker (RW) metric given as
\beq
ds^2=dt^2-a^2(t)\left(\frac{dr^2}{1-kr^2}+r^2d\theta^2+r^2 \sin ^2 \theta d\phi^2 \right) \label{1rw}
\eeq
is widely used isotropic and homogeneous space-time. Here $a(t)$  is  the  scale  factor, $(r,\theta, \phi)$ are spherical polar coordinates  and $k$  is  a  constant indicating the spatial  curvature  of the three space, which can take normalized values   $+1, 0, -1$. When $k = 0$, the three-space is flat.  When $k  =  +1$  and $k  = -1$,  the three-space is of positive and negative  constant curvature; they are known as the closed and open Friedmann models, respectively. 

From the general point of view, the cosmological model can be divided into three parts based on isotropy. The first one is the isotropic model. The second one is known as Locally Rotational Symmetry(LRS)  models in which kinematical quantities are rotationally symmetric about a preferred spatial direction. All observations are rotationally symmetric about this direction at every general point. Third, the obvious one in which observation in a spatial direction differs from observations in other directions is known as \textit{anisotropic} models.

With the help of three independent differential forms defined as $\omega^a=e^a_\alpha dx^\alpha$ where $e^a_\alpha$ are a set of four linearly independent vectors, we can express more general homogeneous space-time metric as
\beq
ds^2=N^2(t)dt^2-\eta_{ab}\omega^a\omega^b,
\eeq
where $\eta_{ab}=e^i_ae_{ib}$ is a symmetric tensor depending on time only.  
Now homogeneity conditions can be expressed as
\beq
C^c_{ab}=\left(\frac{\partial e^c_\alpha}{\partial x^\beta}-\frac{\partial e^c_\beta}{\partial x^\alpha}\right)e^\alpha_ae^\beta_b, \label{1cc}
\eeq 
 where $C^c_{ab}$ are known as structure constants. They are antisymmetric in the lower indices.  All possible homogeneous models can be expressed through the dual of these constants with the help of antisymmetric Levi-Civita tensor $\epsilon_{abc}$, i.e., $C^{ab}=\epsilon^{cda}C^b_{cd}$. Then expression given by (\ref{1cc}) can be written as,\beq
\epsilon_{bcd}C^{cd}C^{ba}=0.  \label{1ecc}
\eeq
We can decompose the  tensor $C^{ab}$ into symmetric and antisymmetric part as
\beq
C^{ab}=n^{ab}+\epsilon^{abc}a_c,
\eeq
where  $\epsilon^{abc}a_c$ is antisymmetric part and $n^{ab}$ is symmetric part. We can redefine symmetric tensor $n_{ab}$ to a diagonal matrix $n_{ab}=diag(n_1,n_2,n_3)$ and similarly  $a_c=(a,0,0)$ without losing the generality.  Then condition given by (\ref{1ecc})  takes the form as 
\beq
an_1=0.
\eeq
\textit{Bianchi} classified various geometric model based on this homogeneity condition. For all possible combinations, $n_1, n_2,n_3$ can assume values $0,1,-1$ and $a\geq 0$. Brief classification of all possible models is given in the table \ref{tab:bianchi}.  One can look at this classification in detail in work done by Bianchi \cite{bianchi}. These models are known as \textit{Bianchi} cosmological models.
\begin{table}
\begin{center}
\begin{tabular}{|c|c|c|c|c|} 
 \hline
 \textbf{Type of Bianchi Model} &  \textbf{ a} &  \textbf{ $n_1$}&  \textbf{ $n_2$}&  \textbf{ $n_3$} \\  
 \hline
I & $ 0$ & $ 0$ & $ 0$ & $ 0$\\ 
II & $ 0$ & $ 1$ & $ 0$ & $ 0$\\
VI & $ 0$ & $ 1$ & $ -1$ & $ 0$\\ 
VII & $ 0$ & $ 1$ & $ 1$ & $ 0$\\ 
\hline
VIII & $ 0$ & $ 1$ & $ 1$ & $ -1$\\ 
IX & $ 0$ & $ 1$ & $ 1$ & $ 1$\\
\hline
III& $ 1$ & $ 0$ & $ 1$ & $ -1$\\ 
IV & $ 1$ & $ 0$ & $ 0$ & $ 1$\\
V & $ 1$ & $ 0$ & $ 0$ & $ 0$\\
$VII_a$ & $ a$ & $ 0$ & $ 1$ & $ 1$\\
$VI_a (a\neq 1)$  & $ a$ & $ 0$ & $ 1$ & $ -1$\\
\hline  
\end{tabular}
\end{center}
\caption{Classification of Bianchi Models.}
\label{tab:bianchi}
\end{table}
\section{The Problem of Non Unitarity}\label{ch:intro6}
Resolving the problem of time, as discussed in section \ref{ch:intro3}, does not solve all the conceptual problems of quantum cosmology. In the standard quantum mechanics, the evolution of wave function, inner product, and conservation of probability is well defined. We need to see whether the Wheeler-DeWitt method of quantization help in defining these for quantum cosmology or not.
 Let's consider the Bianchi-I cosmological model given as 
 \beq \label{1bm1}
 ds^2=N^2(t)dt^2-a^2(t)dx^2-b^2(t)dy^2-c^2(t)dz^2,
 \eeq
 where $a(t),b(t), c(t)$ are respective scale factors along the 3-space directions indicating the anisotropic nature of the model.
 The Lagrangian for the gravity section of the action(\ref{1action3d})  can be given as 
\beq \label{1bml}
 \mathcal{L}_g=-6\frac{e^{3\beta_0}}{N}(\dot{\beta}^2_0-\dot{\beta}^2_+-\dot{\beta}^2_-).
\eeq 
The corresponding Hamiltonian for the gravity sector read as
\beq \label{1hg}
H_g=-\frac{Ne^{-3\beta_0}}{24}(p_0^2-p^2_+-p^2_-),
\eeq 
where $p_i$(for i =0,+,-) are canonical momentas, $p_i=\frac{\partial \mathcal{L}_g}{\partial \dot{\beta}_i}$. 
 With the help of Schutz's formalism, as discussed in section \ref{ch:intro5}, one can find the Hamiltonian for the fluid part, $H_f$ from (\ref{1hf}). Then the total Hamiltonian($H_g+H_f$) for Bianchi I Model takes the form as
 \beq  \label{1bmH}
 H=Ne^{-3\beta_0}\left(-\frac{1}{24}(p_0^2-p^2_+-p^2_-)+e^{3(1-\alpha)\beta_0}p_T\right).
 \eeq
 It is quite apparent from equation (\ref{1bmH}) that the signature of the kinetic term of Hamiltonian is hyperbolic. 
 From equation (\ref{1bmH}) Wheeler-DeWitt equation for Bianchi I Universe can be written as
 \beq
 \left(\frac{\partial^2}{\partial\beta_0^2}-\frac{\partial^2}{\partial\beta_+^2}-\frac{\partial^2}{\partial\beta_-^2}\right)\psi=-24ie^{3(1-\alpha)\beta_0}\frac{\partial\psi}{\partial T}. \label{1bmwdw}
 \eeq
The hermiticity condition of the Hamiltonian requires the wave function $\psi$ of equation (\ref{1bmwdw}) must follow the  boundary condition given  as
\beq \label{1bmbc}
\left(\frac{\partial \psi}{\partial \beta_i}\right)_{\beta_i \rightarrow \pm \infty}=K\left( \psi\right)_{\beta_i \rightarrow \pm \infty} =0.
\eeq
To ascertain whether the wave function for Bianchi I universe obtained from equation (\ref{1bmwdw}) follow the above condition or not, one needs to calculate the expression of the wave function. Alvarenga et al have obtained the explicit expression for the wave function for equation (\ref{1bmwdw}) in work \cite{alvarenga2003}.  The expression for the wave function read as
\begin{multline} \label{1bmwf}
\psi(\beta_i,T)=e^{i(k_+\beta_++k_-\beta_-)}\biggl[c_1J_v\left(\frac{\sqrt{24E}}{3(1-\alpha)/2}e^{3(1-\alpha)\beta_0/2}\right) \\ +c_2J_{-v}\left(\frac{\sqrt{24E}}{3(1-\alpha)/2}e^{3(1-\alpha)\beta_0/2}\right)\biggr]e^{-iET},
\end{multline} 
where $c_1,c_2$ are constant of integration and $J_v$ is Bessel function of order $v$.
By integrating out parameters $k_+,E$ and fixing $k_-=0$, wave packet read as
\beq \label{1bmwp}
\Psi=\frac{1}{B}\sqrt{\frac{\pi}{\gamma}}exp\left[-\frac{e^{3(1-\alpha)\beta_0}}{4B}-\frac{(\beta_++C(\beta_0,B))^2}{4\gamma}\right],
\eeq
where $B=\lambda-i\frac{3(1-\alpha)^2}{32}T$, $C(\beta_0,B)=e^{\beta_0}-\frac{2}{3(1-\alpha)}ln(2B)$ and $\gamma$ and $\lambda$ are constant. It is quite evident that the wave packet given by equation (\ref{1bmwp}) is square integrable and also follows the boundary conditions described by equation (\ref{1bmbc}).

We can obtain the norm $||\Psi||$ from equation (\ref{1bmwp}) as
\beq \label{1bmnorm}
||\Psi ||=\int_{-\infty}^{\infty}\int_{-\infty}^{\infty}e^{3(1-\alpha)\beta_0}\psi^*\psi d\beta_0 d\beta_+=\frac{2\sqrt{2\gamma \pi}}{3(1-\alpha)\lambda}F(T),
\eeq 
where $F(T)=exp\left(\frac{[Img(C(\beta_0,B))]^2}{2\gamma}\right)$, $Img(C(\beta_0,B))=\frac{-2}{3(1-\alpha)}arctan(\frac{-3(1-\alpha)^2T}{32\lambda})$.

From equation (\ref{1bmnorm}), it is quite clear that norm $||\Psi||$ is time dependent; thus, model does not have unitary evolution. 

Time dependent norm also leads to the inequivalence between the \textit{Copenhagen} interpretation\cite{TIPLER1986231} and the \textit{de Broglie-Bohm} interpretation \cite{holland1995quantum}. Even if the Hamiltonian is hermitian, i.e., the eigenvalues are real, the evolution may not be unitary. For unitarity, the Hamiltonian has to be self-adjoint, i.e., $H^\dagger=H$, which means that they act on the same Hilbert space. For a comprehensive discussion on this issue, we refer to the monograph by Reed and Simon\cite{reed}.
 
An operator $A$ can be self adjoint only if the domain of $A^\dagger$ is the same as the domain of $A$. One can calculate the deficiency indices $n_{\pm}$ in order to verify the self-adjointness of the operator $A$. Here $n_{\pm}$ are the dimensions of the linear independent square integrable solutions of the indicial equation given as
\beq \label{1df}
A\phi=\pm i \phi.
\eeq
If $n_+=n_-=0$ then operator $A$ is self-adjoint. Even if $n_{\pm}\neq 0$, a  self-adjoint 'extension' of $A$ is possible if $n_+=n_-$ \cite{reed}. 

Alvarenga et al \cite{alvarenga2003} showed that, for Hamiltonian $H$ given in the form as equation (\ref{1bmH}) which leads to time dependent norm (\ref{1bmnorm}) and the probability is not conserved. Incidentally values of deficiency indices come out as $n_+=0$ and $n_-=1$. Thus $H$ is neither self adjoint, nor it admits any self adjoint extension.

We now discuss the reason for this problem of non-unitarity in the model.  Apparently, the alleged non-unitarity stems from the hyperbolicity of the Hamiltonian (equation \ref{1hg}). The momentum squared terms contribute with both positive and negative signs. FRW models have only one scale factor, so this problem does not arise at all. 

However, it was shown that the clue lies in the operator ordering rather than anything else\cite{spal1}. 

In the case of the Bianchi I model,  Wheeler-DeWitt equation (\ref{1bmwdw}), a particular operator ordering $e^{3(\alpha-1)\beta_0}\frac{\partial^2}{\partial\beta_0^2} $ is used. If one uses the operator ordering as $e^{3(\alpha-1)\beta_0/2}\frac{\partial}{\partial\beta_0}(e^{3(\alpha-1)\beta_0/2}\frac{\partial}{\partial\beta_0})$, then Pal and Banerjee \cite{spal1} show that the Hamiltonian can be written as
\beq
\mathcal{H}_g=\frac{ d^2}{d\chi^2}+\frac{\sigma}{\chi^2},
\eeq
with the help of a new variable $\chi=e^{-\frac{3}{2}(\alpha-1)\beta_0}$. This Hamiltonian is similar to that of the inverse square potential, which is a well studied problem in physics \cite{essin}. Pal and Banerjee show that deficiency indices for $\mathcal{H}_g$, $n_{\pm}=1$; thus, it admits self adjoint extension. Hence the alleged problem of the non-unitarity may be avoided. 
A similar self-adjoint extension is shown for anisotropic models like Bianchi-I, III, IX, Kantowski-Sachs (KS) models in the subsequent work \cite{spal2,spal3,spal4}.

Very recently, Pal and Banerjee \cite{spaljmp} show that a proper ordering can in fact resolve the issue of the alleged non-unitarity. Anyway, operator ordering has a one parameter of U(1) family of self-adjoint extensions. We can have various operator ordering, which leads to a U(1) group of self-adjoint extension; thus, unitarity can be preserved for a bunch of the orderings. Hence the choice of operator ordering is not unique for the unitary evolution of the model. However, the above mentioned particular ordering is a good choice that can preserve the ground state energy, unlike other extensions.  They also emphasize that explicit evaluation of deficiency index and construction of boundary conditions for self-adjoint extension may not be analytically possible in all cases.

The alleged discrepancy between the Copenhagen interpretation and the Bohm-de Broglie interpretation can also be resolved by the proper choice of boundary conditions. However, it may be a non-trivial task to prove it due to the presence of hyperbolicity in the Hamiltonian. Incidentally, this discrepancy is also shown for the isotropic case where the hyperbolicity is absent by Falciano, Pinto-Neto, and Struyve \cite{falc}.

\section{Brans-Dicke Theory}\label{ch:intro7}
The theory of general relativity is based on the equivalence principle. However, Mach's principle \cite{mach}  states that inertial mass is affected by the global distribution of matter; thus, it creates the compatibility issue with general relativity.  Brans and Dicke \cite{brans,dicke} made an attempt to incorporate the Mach's principle in a relativistic theory of the gravity. They introduced a scalar field $\phi=\phi(x^i,t)$, which effectively makes $G$, the Newtonian constant of gravity, vary with the spacetime coordinates.

Brans-Dicke theory of gravity, which is among the widely used modified theory of the gravity. It was believed that the theory reduces to standard general relativity when the coupling constant $\omega \to \infty$. It was later proved that this was not in general correct\cite{nbsen,far1999}. Still, Brans-Dicke theory and some generalization of that find application, particularly in cosmological scenario, such as for resolving the general exit issue of inflation \cite{johri,la1989} or deriving a late time acceleration even without a dark energy\cite{nb2001}.
\subsection{Jordan Frame}
With the assumption that only that Gravitational constant G varies with space-time, action for Brans-Dicke theory can be written as
\beq \label{1bdjf}
A_J=\int d^4x \left[\sqrt{-g}(\phi R + \frac{\omega}{\phi}\partial_\mu \phi \partial^\mu \phi )+ \frac{16\pi}{c^4} L_m\right]
\eeq
where $\phi$ is the scalar field which plays the role analogous to the $G^{-1}$ and $\omega$ is a dimensionless parameter. The second term is the contribution of the Lagrangian density of a scalar field. $L_m$ is the Lagrangian density for matter field.  

This theory, given by equation (\ref{1bdjf}), is also known as the Brans-Dicke theory in Jordan frame for having  a formal connection with Jordan's theory \cite{jordan}.
The field equation for the theory can be written as
\beq
R_{\mu\nu}-\frac{1}{2}g_{\mu\nu}R=\frac{8\pi \phi^{-1}}{c^4}T_{\mu\nu}+\frac{\omega}{\phi^2}(\phi_{,\mu}\phi_{,\nu}-\frac{1}{2}g_{\mu\nu}\phi_{,\delta}\phi^{,\delta})+\frac{1}{\phi}(\phi_{,\mu ;\nu}-g_{\mu\nu}\Box \phi),
\eeq
where $T_{\mu\nu}$ is energy momentum tensor coupled with the variable gravitational parameter $\phi^{-1}$ unlike constant $G$ in the usual Einstein field equation and $\Box$ is covariant d'Alembertian operator. 
 The wave equation for the scalar field $\phi$ is given as
\beq
\Box \phi=\frac{8\pi}{(2\omega+3)c^4}T.
\eeq
\subsection{Einstein Frame}
Later Dicke\cite{dicke} showed the theory in the conventional form, in which the Einstein field equation holds true, with the help of the coordinate dependent transformation of the units of  measure.  Metric tensor transforms as $g_{\mu\nu} \rightarrow \Omega^2 g_{\mu\nu}$ and mass transforms  as $m \rightarrow \Omega m$. Then after affecting the transformation as stated, which is also a conformal transformation, action given by equation (\ref{1bdjf}) takes the form as 
\beq \label{1bdef}
A_E=\int d^4x \sqrt{-\bar{g}}\left[\bar{R}+\frac{2\omega+3}{2}\partial_\mu \bar{\phi} \partial^\nu \bar{\phi} ++ \frac{16\pi}{c^4} \bar{L}_m\right],
\eeq 
where $\bar{R}$ is the Ricci scalar in transformed coordinates, $\bar{\phi}=\ln \phi$ and $\bar{L}_m= L_m/\Omega^4$ is the contribution from matter field. Here we can notice that the scalar field is coupled to the gravity part only in a minimal way, and there is no "interference term" like $\phi R$ as in equation (\ref{1bdjf}). This representation is called as Brans-Dicke Theory in the Einstein frame. If $\Omega^2=\phi$, this frame restores the constancy of the gravitational constant but it effects the rest mass of the particles as it is now a function of the scalar field. This further leads to the loss of the equivalence principle as rest mass is no longer a constant and geodesic equations are no longer valid. In this frame, the field equation can be written as,
\beq
\bar{R}_{\mu\nu}-\frac{1}{2}\bar{g}_{\mu\nu}\bar{R}=\frac{8\pi G_0}{c^4}T_{\mu\nu}+\frac{(2\omega+3)}{2}(\bar{\phi}_{,\mu}\bar{\phi}_{,\nu}-\frac{1}{2}\bar{g}_{\mu\nu}\bar{\phi}_{,\delta}\bar{\phi}^{,\delta}),
\eeq
and wave equation for the scalar filed is
\beq
\Box \bar{\phi}=\frac{8\pi}{(2\omega+3)c^4}T.
\eeq

Now the bigger question arises: which frame should be used? Physicists are divided in order to answer this question. This topic is widely debated, and it can be categorized. There are authors\cite{Buchmuller:1988cj,Holman:1990wq,Campbell:1990de,Shapiro:1995kt,Kaloper:1997sh} argue that the two frames are physically equivalent. In works \cite{Jakubiec:1988ef,Hu:1993jc,Hwang:1990re,Suzuki:1990en,Capozziello:1996xg}, physicists consider these frames are physically non-equivalent for various reasons. The other group of authors \cite{Damour:1990eh,Holman:1990hg,Wu:1991dt,Steinhardt:1994vs,Barros:1997qt} regards only Jordan frame as a physical frame, but they show that the Einstein frame can also be used for mathematical convenience.   Many physicists also believe that the Einstein frame is the only physical frame\cite{Gibbons:1987ps,Cho:1987xy,Deruelle:1991zi,cho1992,Cotsakis:1994eb,Fujii:1997wa}. Detail summary can be found in review work by Faraoni, Gunzig, and Nardone\cite{FGN}. For quantum aspects of the equivalence, we refer to the discussion by Almedia et al\cite{almeida2017quantum, fabris2018}.
  
\section{Outline of the present work}\label{ch:intro9}

The purpose of the thesis is two fold. One is to look at some anisotropic models in connection with the possibility of unitary evolutions of quantized models in the Wheeler-DeWitt formulation. Chapters \ref{ch2:1}, \ref{ch3:1} and \ref{ch4:1} are devoted to that. The second purpose is to use Wheeler-DeWitt formulation to find the resolution of the equivalence of Jordan and Einstein frames at the quantum level. Chapters \ref{ch5:1} and \ref{ch6:1} deal with that.
 
 In chapter \ref{ch2:1}, we quantize the Bianchi II and Bianchi VI cosmological models, which are anisotropic models with variable spatial curvature, using the Wheeler-DeWitt method of quantization. As time is itself a coordinate in a relativistic theory, dynamical variables related to fluid have been used as \textit {``time''} following Schutz's formalism. We show unitary evolution is indeed possible for these models as their respective Hamiltonians admit self-adjoint extensions. 
 
In chapter \ref{ch3:1}, we work on extending the unitary evolution in higher dimensional anisotropic quantum cosmological models. We discuss the Wheeler-DeWitt quantization scheme for the model with a perfect fluid in the presence of a massless scalar field. We identify the time parameter using a generalization of Schutz's formalism and find the wave packet of the universe following standard Wheeler-DeWitt methodology of quantum cosmology. We establish the unitary evolution for the model. We also calculate the expectation values of scale factors and volume elements for different dimensions, which show that the quantized model escapes the singularity and supports the bouncing universe solutions.

In order to achieve the unitary evolution of cosmological models, what are the price one has to pay for the self-adjoint extension, do we lose symmetries like Noether symmetry and scale-invariance? This has been discussed for the Bianchi-I cosmological model in chapter \ref{ch4:1}.  

There has been a long standing debate regarding the equivalence of  Jordan and Einstein frames in literature. We quantize the Brans-Dicke theory in both the frame using dynamic variable related to the scalar field as time parameter and then address the question of equivalence of these two frames at the quantum level for the isotropic cosmological model, i.e., FRW in chapter \ref{ch5:1}. The obtained expressions for wave packets in both  the frames and show the equivalence of the frames at the quantum level.

In chapter \ref{ch6:1}, we use a similar method of the quantization, i.e., Wheeler-DeWitt, with dynamic variables related to the scalar field as the time for the various anisotropic model, i.e., Bianchi I, V, IX and LRS Bianchi -I and Kantowski-Sachs models in both the frames. We try to generalize the equivalence of both frames at the quantum level for all these models.

\chapter[Anisotropic cosmological models with variable spatial curvature]{Anisotropic cosmological models with variable spatial curvature\footnote{ The work illustrated in this chapter is published; \textbf{S. Pandey} and N. Banerjee, Phys. Scr. \textbf{91}, 115001(2016).}}\label{ch2:1}

Quantum cosmology has its own motivation, such as looking for a resolution of the problem of singularity at the birth of the universe. The basic framework for quantum cosmology is provided by the Wheeler-DeWitt  equation\cite{dewitt1967quantum, wheeler1987superspace, misner1969quantum}.
The Wheeler-DeWitt formulation actually has a very general appeal, the approach is very similar to the usual practice in standard quantum physics. From the classical Lagrangian, the momenta corresponding to the identified coordinates are found out so as to write the Hamiltonian, the variables are then promoted to operators (usually in the coordinate representation), and the relevant Schrodinger like equations are found out which govern the system. In some cases, the Wheeler-DeWitt equation has a straightforward analogue with some situations in other branches of physics. For instance, one anisotropic quantum cosmological model, namely the Bianchi-I model eventually reduces to the standard quantum mechanical problem with an inverse square potential\cite{spal1} which has many applications in other branches of physics\cite{gopal}. 

One major problem of quantum cosmology is that the quantization of anisotropic models are believed to give rise to a non-unitary evolution of the wave function resulting in a non-conservation of probability. One may note that this non-unitary evolution often gets undetected in the absence of a properly oriented scalar time parameter in the quantization methodology.\cite{lidsey1995wave, pinto2000quantum}.  
 A novel idea about the identification of time through the evolution of a fluid present in the model appeared to work very well. The method, where the fluid variables are endowed with dynamical degrees of freedom through some thermodynamic potentials\cite{schutz1970perfect, schutz1971hamiltonian}, was suggested by Lapchinskii and Rubakov\cite{Lapchinskii}. It has been shown that the time parameter that emerges out of the fluid evolution has the required monotonicity as well as the correct orientation\cite{spal1}. This Schutz formalism is now very widely used in quantizing cosmological models\cite{alvarenga1998dynamical, alvarenga2002quantum, alvarenga2003, Majumder2013, almeida2015quantum,spal1, spal2, spal3} and is described in detail in Section \ref{ch:intro4}.\\

Until very recently, the non-conservation of probability in anisotropic models had almost been generally accepted as a pathology, and had been ascribed to the hyperbolicity of the Hamiltonian\cite{alvarenga2003}. Not that the anisotropic models 
are of utmost importance so far as the observed universe is concerned, but this feature of non-unitarity renders the quantization scheme vulnerable. It should also be mentioned that observations do indicate anisotropy in the Cosmic Microwave Background, which however is quite compatible with an isotropic universe statistically. This fluctuations are essentially local effects, consistent with the requirements for the structure formation. \\

There has now been a new turn in this picture. It is clearly shown by Pal and Banerjee\cite{spal1, spal2} that the said non-unitarity can actually be attributed to either an ordering of operators or to a bad choice of variables. With a suitable ordering, examples of unitary evolution were exhibited in Bianchi I, V and IX models. However, even a few examples are good enough to indicate that the problem is not actually pathological and can be cured. Very recently, an example of a unitary evolution for a Kanotowki-Sachs model has been given by Pal and Banerjee\cite{spal3}. It was also shown by Pal\cite{spal4} that this unitarity is achieved not at the cost of anisotropy itself. \\

Except for the Kantowski-Sachs cosmology, all other examples of the anisotropic Bianchi models stated have one unifying feature, they are all of the constant spatial curvature. The motivation for the present work is to show that the possibility of a  self adjoint extension and hence a unitary evolution is not a characteristic of models with a constant spatial curvature, this is in fact more general and can be extended to models with variable curvature of spatial hypersurfaces as well. Two specific examples, namely Bianchi II and VI are dealt with in this chapter. 
\section{The formalism and Bianchi VI models}

We start with the standard Einstein-Hilbert action for gravity along with a perfect fluid given by 

\begin{equation}
\label{c1action}
{\mathcal A} = \int_M d^4x\sqrt{-g}R +\int_M d^4x\sqrt{-g}P,
\end{equation}

where $R$ is the Ricci Scalar, $g$ is the determinant of the metric and $P$ is the pressure of the ideal fluid related to density $\rho$ by an equation of state ($\mathcal{P}=\alpha\rho$). The first  term corresponds to the gravity sector and the second term is due to the matter sector. Here we have ignored the contributions from boundary as it would not contribute to the variation. The units are so chosen that $16\pi G =1$.\\

A Bianchi VI model is given by the metric
\begin{eqnarray}
ds^2 = n^2(t)dt^2-a^2(t)dx^2-e^{-mx}b^2(t)dy^2-e^xc^2(t)dz^2,
\label{c1metric-6}
\end{eqnarray}

where the lapse function $n$ and $a, b, c$ are functions of time $t$ and $m$ is a constant. \\

From the metric given above, we can write the Ricci Scalar as
\begin{multline}
\label{c1ricci-6}
\sqrt{-g}R= e^{\frac{(1-m)x}{2}} \bigg[\frac{d}{dt}[\frac{2}{n}(\dot{a}bc +\dot{b}ca+a\dot{c}b)] -\frac{2}{n}[\dot{a}\dot{b}c +\dot{b}\dot{c}a+\dot{c}\dot{a}b \\ +\frac{n^2bc}{4a}(m^2-m+1)]\bigg].
\end{multline}

Using this, we can find the  action for the gravity sector from  equation (\ref{c1action}) which is given as
\begin{equation}
\label{c1action-grav}
{\mathcal A}_g=\int dt \bigg[-\frac{2}{n}[\dot{a}\dot{b}c+\dot{b}\dot{c}a+\dot{c}\dot{a}b+\frac{n^2bc}{4a}(m^2-m+1)]\bigg],
\end{equation}
where an overhead dot indicates a derivative with respect to time. \\

Now  we make a set of transformation of variables as 
\begin{eqnarray}
a(t)=e^{\beta_0}, \\ 
b(t)=e^{\beta_0+\sqrt{3}(\beta_+-\beta_-)}, \\
c(t)=e^{\beta_0-\sqrt{3}(\beta_+-\beta_-)}.
\end{eqnarray}
This introduces a constraint $a^2=bc$, but the model still remains Bianchi Type VI without any loss of the typical characteristics of the model. Such type of transformation of variables has been extensively used in the literature\cite{spal1, Majumder2013, alvarenga2003}. One can now write the Lagrangian density of the gravity sector as 
\begin{equation}
{\mathcal L}_g = -6\frac{e^{3\beta_0}}{n}[\dot{\beta_0^2}-(\dot{\beta_+}-\dot{\beta_-})^2 +\frac{e^{-2\beta_0}n^2(m^2-m+1)}{12}]. \label{c17}
\end{equation}

Here $\beta_0$ ,$\beta_+$ and $\beta_-$ has been treated as coordinates. So corresponding Canonical momentum will be $p_0$, $p_+$ and $p_-$ where $p_{i} = \frac{\partial {\mathcal L}_g}{\partial \dot{\beta_{i}}}$. It is easy to check that one has $p_+ =-p_-$. Hence we can write the corresponding Hamiltonian as
\begin{equation}
{\mathcal H}_g=-n e^{-3\beta_0}[\frac{1}{24}(p_0^2-p_+^2-12(m^2-m+1)e^{4\beta_0})]. \label{c18}
\end{equation}

With the widely used technique, developed by Lapchinskii and Rubakov\cite{Lapchinskii} by using the Schutz's formalism  as discussed in section \ref{ch:intro4}, the action the fluid sector can be written as
\begin{equation}
\label{c1action-matter}
{\mathcal A}_f =\int dt {\mathcal L}_{f}\\ = \int dt \left[n^{-\frac{1}{\alpha}}e^{3\beta_{0}}\frac{\alpha}{\left(1+\alpha\right)^{1+\frac{1}{\alpha}}}\left(\dot{\epsilon}+\theta\dot{S}\right)^{1+\frac{1}{\alpha}}e^{-\frac{S}{\alpha}}\right].
\end{equation}

Here $\epsilon, \theta, S$ are thermodynamic potentials. A constant spatial volume factor $V$ comes out of the integral in both of (\ref{c1action-grav}) and (\ref{c1action-matter}). This $V$ is inconsequential and can be absorbed in the subsequent variational principle. With a canonically transformed set of variables $T,\epsilon^{\prime}$ in place of $S, \epsilon$, one can finally write down the Hamiltonian for the fluid sector as

\begin{equation}
{H}_f = n  e^{-3\beta_0}e^{3(1-\alpha)\beta_0}p_T.
\label{c143}
\end{equation}

The canonical transformation is given by the set of equations

\begin{eqnarray}\label{c1canonical}
T&=&-p_{S}\exp(-S)p_{\epsilon}^{-\alpha -1},\\
p_{T}&=&p_{\epsilon}^{\alpha+1}\exp(S),\\
\epsilon^{\prime}&=&\epsilon+\left(\alpha+1\right)\frac{p_{S}}{p_{\epsilon}},\\
p_{\epsilon}^{\prime}&=&p_{\epsilon},
\end{eqnarray}

This method and the canonical nature of the transformation are comprehensively discussed in reference \cite{spal1}. \\

The net or the super Hamiltonian is
\begin{equation}
H= H_g + H_f = -\frac{ne^{-3\beta_0}}{24}[p_0^2-p_+^2-12(m^2-m+1)e^{4\beta_0}-e^{3(1-\alpha)\beta_0}p_T] .
\label{c144}
\end{equation}
Using the Hamiltonian constraint $H=0$, which can be obtained by varying the action ${\mathcal A}_{g}+ {\mathcal A}_{f}$ with respect to the lapse function $n$, one  can write the Wheeler-DeWitt equation as
\begin{equation}
[e^{3(\alpha-1)\beta_0}\frac{\partial^2}{\partial \beta_0^2}-e^{3(\alpha-1)\beta_0}\frac{\partial^2}{\partial \beta_+^2}+12(m^2 - m + 1)e^{(3\alpha+1)\beta_0}]\psi =24i\frac{\partial}{\partial T}\psi.
\label{c145}
\end{equation}
This equation is obtained after we promote the momenta to the corresponding operators given by $p_{i}=-i\frac{\partial}{\partial {\beta}_{i}}$ in the units of $\hbar=1$. \\

It is interesting to note that for a particular value of $m=m_0$ where $m_0$ is a root of equation $m^2-m+1 = 0$, the spatial curvature vanishes and the equation (\ref{c145}) reduces to the corresponding equation for a Bianchi Type I model\cite{spal1}. We shall discuss the solution of the Wheeler-DeWitt equation in two different cases, namely $\alpha = 1$ and $\alpha \neq 1$. \\ 

\subsection{Stiff fluid: $\alpha = 1$}

For a stiff fluid ($P=\rho$), the equation (\ref{c145}) becomes simple and easily separable. It looks like
\begin{equation}
\bigg[\frac{\partial^2}{\partial \beta_0^2}-\frac{\partial^2}{\partial \beta_+^2}+12(m^2-m+1)e^{4\beta_0}\bigg]\psi  =24i\frac{\partial}{\partial T}\psi .
\label{c119}
\end{equation} 

With the separation ansatz 

\begin{equation}
\psi = e^{i2 k_+\beta_+}\phi(\beta_0)e^{-iET},
 \label{c120}
\end{equation}
one can write 
 \begin{equation}
\frac{\partial^2 \phi}{\partial \beta_0^2}+(4k_+^2-24E+4N^2 e^{4\beta_0})\phi=0,
\end{equation}
 where $N^2=3(m^2-m+1)$. After making the change in variable as $q = N e^{2\beta_0}$, above equation can be written as
 \begin{equation}
q^2\frac{\partial^2 \phi}{\partial q^2}+q\frac{\partial \phi}{\partial q}+[q^2 - (6E-k_+^2)]\phi=0.
\end{equation}
Solution of this equation can be written in terms of Bessel's functions as 
\begin{equation}
\phi(q) = J_{\nu} (q)  \label{c1phi_q},
\end{equation}

where $\nu =\sqrt{6E-k_+^2}$. Now for the construction of the wave packet, we need to fix $\nu$. If we take $\epsilon= -\nu^2 =k_+^2-6E$ then wave packet can have the following expression 
\begin{equation}
\Psi = \Phi (q) \zeta(\beta_+) e^{i\epsilon T/6}.
\end{equation}
where \begin{equation}
\zeta(\beta_+)=\int dk_+ e^{-(k_+ -k_{+0})^2} e^{i (2k_+\beta_+ - \frac{k_+^2}{6} T)}
\end{equation}

The norm indeed comes out to be positive and finite (for the details of the calculations, we refer to work of Pal and Banerjee \cite{spal3}). Thus one indeed has a unitary time evolution. \\

\subsection{General perfect fluid: $\alpha \neq 1$}

Now we shall take the more complicated case of $\alpha \neq 1$ and try to solve the Wheeler-DeWitt equation (\ref{c145}).
We use a specific type of operator ordering with which equation (\ref{c145}) takes the form
\begin{equation}
\bigg[e^{\frac{3}{2} (\alpha-1)\beta_0}\frac{\partial}{\partial\beta_0}e^{\frac{3}{2} (\alpha-1)\beta_0}\frac{\partial}{\partial\beta_0}-e^{3(\alpha-1)\beta_0}\frac{\partial^2}{\partial\beta_+^2}
+
12(m^2-m+1)e^{(3\alpha+1)\beta_0}\bigg]\Psi=24i\frac{\partial}{\partial T}\Psi. \label{c125} 
\end{equation} 

Now with the standard separation of variable as,
\begin{equation}
\Psi(\beta_0,\beta_+ ,T) =\phi(\beta_0) e^{ik_+\beta_+}e^{-iET},
\end{equation}
the equation for $\phi$ becomes
\begin{equation}
\bigg[e^{\frac{3}{2} (\alpha-1)\beta_0}\frac{\partial}{\partial\beta_0}e^{\frac{3}{2} (\alpha-1)\beta_0}\frac{\partial}{\partial\beta_0}+e^{3(\alpha-1)\beta_0}k_+^2+12(m^2-m+1)e^{(3\alpha+1)\beta_0}-24E\bigg]\phi=0. \label{c126}
\end{equation}
For $\alpha \neq 1$ we make a transformation of variable as
\begin{equation}
\chi =e^{-\frac{3}{2} (\alpha-1)\beta_0},
\end{equation}
and write equation (\ref{c126}) as
\begin{equation}
\frac{9}{4}(1-\alpha)^2\frac{\partial^2\phi}{\partial \chi^2}+\frac{k_+^2}{\chi^2}\phi +12(m^2-m+1)\chi^{\frac{2(3\alpha+1)}{3(1-\alpha)}}\phi-24E\phi =0. \label{c127}
\end{equation}
We define some parameters as
\begin{eqnarray}
\sigma =\frac{4k_+^2}{9(1-\alpha)^2}, \\  
E' = \frac{32}{3(1-\alpha)^2}E,\\
M^2 =\frac{16(m^2-m+1)}{3(1-\alpha)^2}. \label{c128}
\end{eqnarray}
Equation (\ref{c127}) can now be written as
\begin{equation}
-\frac{\partial^2\phi}{\partial \chi^2}-\frac{\sigma^2}{\chi^2}\phi -M^2\chi^{\frac{2(3\alpha+1)}{3(1-\alpha)}}\phi=-E'\phi. \label{c129}
\end{equation}
Above equation can be compared to $-{\mathcal H}_g=-\frac{d^2}{d\chi^2}+V(\chi)$ with $V(\chi)=-\frac{\sigma^2}{\chi^2} -M^2\chi^{\frac{2(3\alpha+1)}{3(1-\alpha)}}$ which is a continuous and real valued function on the half line. One can show that the Hamiltonian $H_g$ admits self-adjoint extension as ${\mathcal H}_g$ has equal deficiency indices. We can refer to the text of Reed  and Simon\cite{reed} for a systematic and detailed description of the self-adjoint extension. \\
So it can be said that for perfect fluid with $\alpha \neq 1$, Bianchi VI quantum models do admit  a unitary evolution.\\

\subsection{$\alpha=-\frac{1}{3}$}
We take a specific choice, where $\rho+3P =0$, as an example. This equation of state will make equation (\ref{c129}) much simpler. 
With $\alpha=-1/3$, the term $-M^2\chi^{\frac{2(3\alpha+1)}{3(1-\alpha)}}$ becomes a constant ($M^{2}$). Equation (\ref{c129}) becomes

\begin{equation}
-\frac{\partial^2\phi}{\partial \chi^2}-\frac{\sigma^2}{\chi^2}\phi =-(E'-M^2)\phi, \label{c130}
\end{equation}
which is in fact a well known Schrodinger equation of a particle with mass $m=1/2$ in an attractive inverse square potential. Solution to above can be given as,
\begin{eqnarray}
\phi_a(\chi)=\sqrt{\chi}[AH_{i\beta}^{(2)}(\lambda \chi)+BH_{i\beta}^{(1)}(\lambda \chi)], \\
\phi_b(\chi)=\sqrt{\chi}[AH_{\alpha}^{(2)}(\lambda \chi)+BH_{\alpha}^{(1)}(\lambda \chi)], \label{c131}
\end{eqnarray}
for  $\sigma >1/4$ and $\sigma < 1/4$ and $\beta = \sqrt{\sigma-1/4}$ and $\beta = \sqrt{1/4-\sigma}$ respectively. Here both $\alpha$ and $\beta$ are real numbers and in both cases the energy spectra is given as
\begin{equation}
 E'=M^2-\lambda^2.
\end{equation}

For ${\mathcal H}_g=\frac{d^2}{d\chi^2}+\frac{\sigma^2}{\chi^2}$, the value of deficiency index $n_{\pm}$, which is the number of linearly independent solutions for equation  ${\mathcal H}_g\phi_{\pm}=\pm i\phi_{\mp}$,  comes out to be $n_+=n_-=1$. It is always possible to have self adjoint extension of the Hamiltonian having  equal deficiency i.e. indices $n_+=n_-$. For an inverse square potential, the method is described in detail by Essin and Griffiths \cite{essin}. Using the asymptotic expression for $\phi_a$ and $\phi_b$, self-adjoint extension   guarantees that  $|B/A|$ takes a value so as to conserve probability and make the model unitarity. 
\section{Bianchi II models:}

Bianchi Type II model is given by the line element
\begin{equation}
ds^2=dt^2-a^2(t)dr^2-b^2(t)d\theta^2-[a^2(t) \theta^2 +b^2(t)]d\phi^2+2a^2(t)\theta dr d\phi.
\label{c151} 
\end{equation}

The calculation in this case is a bit more involved for the presence of the non-diagonal terms in the metric. \\

The Ricci scalar $R$ in this case is given by
\begin{equation}
R = -\frac{a^2}{2 b^4} - \frac{4\dot{a}\dot{b}}{a b} -\frac{2{\dot{b}}^2}{b^2} -\frac{2\ddot{a}}{a} -\frac{4\ddot{b}}{b}.
\end{equation}

If we define a new variable $\beta=a b$ as prescribed in \cite{alvarenga2003}, then Lagrangian density for gravity sector looks like
\begin{eqnarray}
 {\mathcal L}_g =\frac{2\beta^2\dot{a}^2}{a^3}-\frac{2\dot{\beta}^2}{a}-\frac{a^5}{2\beta^2}, 
\label{c152} 
\end{eqnarray}
 and the corresponding Hamiltonian density for gravity sector can be written as
 \begin{equation}
H_g=\frac{a^3p_a^2}{8\beta^2}-\frac{a}{8}p_{\beta}^2+\frac{a^5}{2\beta^2} . 
\label{c153}
\end{equation}
Using Schutz's formalism and proper identification of time as we did before, the Hamiltonian density for fluid sector can be written as 
\begin{equation}
H_f =  a^{\alpha}\beta^{-2\alpha}p_T.
\label{c154}
\end{equation}
The super Hamiltonian can now be written in following form
\begin{equation}
H = H_g + H_f = \frac{a^3p_a^2}{8\beta^2}-\frac{a}{8}p_{\beta}^2+\frac{a^5}{2\beta^2} +a^{\alpha}\beta^{-2\alpha}p_T .
\label{c155}
\end{equation}

As an example we take up the case of a stiff fluid given by $\alpha=1$. 

After promoting the momenta by operators as usual, the Wheeler-DeWitt equation  $H\Psi=0$ takes following form
\begin{equation}
-\frac{a^2}{8}\frac{\partial^2 \psi}{\partial a^2}+\frac{\beta^2}{8}\frac{\partial^2 \Psi}{\partial \beta^2}+\frac{a^4}{2}\Psi = i \frac{\partial \Psi}{\partial T}.
\label{c157}
\end{equation}
Using a separation of variables
\begin{equation}
\Psi =e^{-iET}\phi(a)\psi(\beta),
\end{equation}

we get following equations for $\psi$ and $\phi$ respectively
\begin{eqnarray}
-\frac{d^2\psi}{d\beta^2}+\frac{8k}{\beta^2}\psi=0, \label{c158}\\
a^2\frac{d^2\phi}{da^2}-4a^4\phi-8(k-E)\phi=0. \label{c159}
\end{eqnarray}

With $\phi=\frac{\phi_0}{\sqrt{a}}$ and $\chi=a^2$, last equation can be written as
\begin{equation}
-\frac{d^2\phi_0}{d\chi^2}-\frac{\sigma}{\chi^2}\phi_0=-\phi_0 ,\label{c160}
\end{equation}
where $\sigma=[\frac{3}{16}-2(k-E)].$ Equations (\ref{c158}) and (\ref{c160}) are the governing equations for Bianchi Type II with a stiff fluid. \\

Equations for both $\psi$ and $\phi$ can be mapped to a Schrodinger equation for a particle in an inverse square potential. In order to get a solution we actually have ensure an attractive regime, which requires $k\leq 0$ ,  $E \leq k-3/32$. We see that both the equations are that for inverse square potentials, and thus a self-adjoint extension is possible. 

\section{A note on spatial curvature}

It has already been mentioned that the difference between the present examples of Bianchi VI and Bianchi II models on one hand, and most of the models discussed earlier on the other, is the fact that the present models have variable spatial curvature as opposed to most the models discussed in connection with the quantization of cosmological models according to the Wheeler-DeWitt scheme. In a (3+1) decomposition of the space-time metric, one can calculate the Ricci curvature ($^{3}R$) of the three dimensional space section, embedded in a four dimensional space-time. Both Bianchi VI and II will have $^{3}R$ which vary with time. For example, if we take the Bianchi VI metric as an example in a more general form, than that used here (equation (\ref{c1metric-6})), given by

\begin{equation}
\label{c1metric-6-gen}
 ds^2 = n^2(t)dt^2-a^2(t)dx^2-e^{-mx}b^2(t)dy^2-e^{lx}c^2(t)dz^2,
\end{equation}

the 3-space Ricci curvature looks like 

\begin{equation}
\label{c1ricci-curv}
^{3}R = \frac{m^{2} - ml + l^{2}}{2 a^{2}(t)},
\end{equation}
 which indeed is a function of the cosmic time $t$ through $a$. In the case of the metric (\ref{c1metric-6}), this becomes

\begin{equation}
\label{c1ricci-curv1}
^{3}R = \frac{m^{2} - m + 1}{2a^{2}(t)}.
\end{equation}

The most talked about anisotropic models, like Bianchi I, V and IX all have costant $^{3}R$. For example, if we put $m=l=0$ in the metric (\ref{c1metric-6-gen}), we get a Bianchi I metric, and for this choice, equation (\ref{c1ricci-curv}) clearly shows that $^{3}R = 0$. For some more information regarding the spatial curvature, we refer to the recent work by Akarsu and Kilinc\cite{akarsu}.\\

\section{Discussion and conclusion}

The work discussed in this chapter deals with two examples of anisotropic quantum cosmological models with varying spatial curvature. We show that there is indeed a possibility of finding unitary evolution of the system. The earlier work on anisotropic models with constant spatial curvature\cite{spal1, spal2} disproved the belief that anisotropic quantum cosmologies generically suffer from a pathology of non-unitarity.  The present work now strongly drives home the fact that this feature is not at all a charactristic of models with constant spatial curvature. It was also shown before that the unitarity is not achieved at the cost of anisotropy itself\cite{spal4}. One can now indeed work with quantum cosmologies far more confidently, as there is actually no built-in generic non-conservation of probability in the models. \\

Very recently it has been shown that in fact all homogeneous models, isotropic or anisotropic, quite generally have a self-adjoint extension\cite{spaljmp}, although the extension is not unique. The present work gives two more examples, and consolidates the result proved in reference \cite{spaljmp}. The examples chosen indeed have physical implications. It has been shown very recently that a Bianchi VI model plays an important role in producing anisotropic inflation\cite{kao2011}. We also refer to the work of Barrow\cite{barrow1984} for various cosmological implications of Bianchi type VI models. Bianchi II models, on the other hand, are instrumental in understanding the Belinskii, Khalatnikov, Lifshitz conjecture in the discussion of spacelike singularities\cite{bkl1970, bkl}. \\

Thus the standard canonical quantization of cosmological models via Wheeler-DeWitt equation still proves to be useful in the absence of a more general quantum theory of gravity.
\begin{titlepage}
\newpage
\clearpage
\newpage
\mbox{~}
\clearpage
\newpage
\end{titlepage}
\chapter[Quantization of n-dimensional cosmological models]{Quantization of n-dimensional cosmological models\footnote{ The work illustrated in this chapter is published; \textbf{S. Pandey} , European Physical Journal C, \textbf{79}, 487(2019).}}\label{ch3:1}

Higher dimensional models had been investigated widely in the past in order to find a theory to unify gravity with other fundamental forces of physics. It started with Kaluza and Klein's assertion that a fifth dimension in general relativity will unify gravity with the electromagnetic field\cite{kalu,klein,klein1}. Further motivation of resorting to higher dimensional models came from the expectation of unifying gravity with non-Abelian gauge fields\cite{cho1975higher,cho1975non}. Later, spacetime having more than four dimensions were motivated by 10-dimensional superstring theory and 11-dimensional supergravity theory\cite{ss,sg}.

In Chapter \ref{ch2:1}, we use the Wheeler-DeWitt quantization method to quantize anisotropic models, where the evolution of the cosmic fluid is identified as a time parameter using Schutz formalism. In cosmological models related to Brans-Dicke theory\cite{brans,dicke}, the evolution of the scalar field have been used as a time parameter in order to quantize the same without adding any additional matter\cite{vakili2012scalar,sachin1,sachin2}. Brans-Dicke theory of gravity has also been quantized using Schutz's formalism\cite{almeida2017quantum,spal5}. Similar formalism have been used by Khodadi  {\textit et al}  in scalar-energy dependent metric cosmology\cite{khodadi2016classical}.  
So long as a method gives rise to an oriented scalar time parameter, it is quite an effective method.

In a very recent work, Alves-Junior {\textit et al}\cite{ndaniso} have shown the canonical quantization of n-dimensional anisotropic model coupled with a massless scalar field in the absence of any fluid where they have emphasized on natural identification of scalar field as time parameter for the evolution of quantum variables.\\

In this chapter, we try to investigate the quantum cosmological solutions of an n-dimensional anisotropic model in which massless scalar field is minimally coupled with gravity in presence of a barotropic perfect fluid($P=\alpha \rho$). The idea is to generalize the work of Alves-Junior { et al} to include a fluid, as well as to generalize the recent work on anisotropic cosmologies by Pal and Banerjee\cite{spal1,spal2} to higher dimensions. Keeping in mind the importance of various aspects of higher dimensions, it is quite relevant to ask the questions like whether singularity-free n-dimensional quantum cosmological models may be realized. Also, this investigation should confirm if the general result, given by Pal and Banerjee\cite{spaljmp}, that it is always possible to find a self-adjoint extension of homogeneous quantum cosmologies in four dimensions, is applicable to higher dimensions also.
\section{An n-dimensional cosmological model}
\subsection{The action}
We start with the Hilbert-Einstein action in n-dimension with a minimally coupled scalar field in the presence of a fluid,
\begin{equation}\label{c2action}
A=\int d^nx \sqrt{-g}(R+\omega g^{\mu\nu}\phi_{,\mu}\phi_{,\nu})+\int d^nx \sqrt{-g}P, 
\end{equation}
where $\omega$ is a dimensionless parameter and $\phi$ is the massless scalar field. The last term of equation (\ref{c2action}) represents the matter contribution coming from fluid pressure $P$ related with density $\rho$ by the equation of state $P=\alpha \rho$.
The metric for an n dimensional ($n>4$) spatially homogeneous but anisotropic cosmology is chosen as  
\begin{equation}\label{c2met}
ds^2=N(t)^2dt^2-a(t)^2(dx^2+dy^2+dz^2)-b(t)^2\sum^{n-4}_{i=1} dl_i^2,
\end{equation}
where $N(t)$ is the lapse function, $a(t)$ is our good old scale factor used in flat FLRW cosmology and $b(t)$ is scale factor coming from remaining $(n-4)$ dimensions. So the usual 3-space is isotropic, but the extra dimensions, though isotropic in itself, is anisotropic with respect to the usual 3-space section. \\
The action for scalar gravity part can be written in a reduced form for metric given by (\ref{c2met}) as 
\begin{multline} \label{c2gaction}
A_g=V_0\int dt \frac{6}{N}\biggl[ -\dot{a}^2ab^{n-4}-(n-4)\dot{b}\dot{a}a^2b^{n-5}\\-\frac{(n-4)(n-5)}{6}\dot{b}^2a^3b^{n-6}  +\frac{\omega}{6}a^3b^{n-4}\dot{\phi}^2\biggr],
\end{multline} 
where an over-dot represents a derivative with respect to coordinate $t$ and $V_0$ denotes $(n-1)$ dimensional volume. We have ignored the surface terms as they do not contribute to the field equations.\\
From (\ref{c2gaction}), we can write the Lagrangian for the gravity sector as, 
\begin{multline} \label{c2lg}
L_g= -\frac{6}{N}\dot{a}^2ab^{n-4}-\frac{6(n-4)}{N}\dot{b}\dot{a}a^2b^{n-5}\\-\frac{(n-4)(n-5)}{N}\dot{b}^2a^3b^{n-6}+\frac{\omega}{N}a^3b^{n-4}\dot{\phi}^2.
\end{multline}
Schutz \cite{schutz1970perfect,schutz1971hamiltonian} showed that four dimensional velocity vector of a perfect fluid can be written using six thermodynamic quantities as,
\begin{equation} \label{c24v}
U_\nu=\frac{1}{\mu}\left(\epsilon_{,\nu}+\zeta \beta_{,\nu}+\theta S_{,\nu} \right),
\end{equation} 
where $\mu$ is specific enthalpy, $S$ is specific entropy, $\zeta$ and $\beta$ are potentials connected with rotation and $\epsilon$ and $\theta$ are potentials with no clear physical meaning. For n dimensional velcoity vector, we can consider additional potential terms\cite{ndfrw} like $\zeta$ and $\beta$  in equaution(\ref{c24v}) as
\begin{equation} \label{c2nv}
U_\nu=\frac{1}{\mu}\left(\epsilon_{,\nu}+\zeta \beta_{,\nu}+\theta S_{,\nu} +\sum^{n-4}_{i=1}\zeta_i \beta_{i,\nu}\right).
\end{equation}  Since the potentials $\zeta$, $\zeta_i$, $\beta$ and$\beta_i$  are related to rotation, we can neglect them in the present case as there is no vorticity in the spacetime given by equation  (\ref{c2met}). Following the widely used technique, developed by Lapchinskii and Rubakov \cite{Lapchinskii} by writing the fluid pressure in terms of potentials given in equation (\ref{c2nv}), the fluid sector action given in equation (\ref{c2action}) can be written as
\begin{equation}
 A_f=V_0\int dt \biggl[ N^{-1/\alpha} a^3 b^{n-4}\frac{\alpha}{(\alpha +1)^{1/\alpha +1}}(\dot{\epsilon}+\theta\dot{S})^{1/\alpha +1}  exp(-S/\alpha)\biggr].
 \end{equation}
From above equation, Lagrangian for fluid sector can be written as
 \begin{equation} \label{c2lf}
 L_f= N^{-1/\alpha} a^3 b^{n-4}\frac{\alpha}{(\alpha +1)^{1/\alpha +1}}(\dot{\epsilon}+\theta\dot{S})^{1/\alpha +1} exp(-S/\alpha).
 \end{equation}
 \subsection{Hamiltonian of the model}  
Following standard procedure \cite{gold}, Hamiltonian for gravity sector from corresponding Lagrangian (\ref{c2lg}) can be written as
\begin{multline}\label{c2hg}
H_g=\frac{N}{(n-2)ab^{n-6}}\biggl[ \frac{(n-5)}{12}\frac{p_a^2}{b^2}+\frac{1}{2(n-4)}\frac{p_b^2}{a^2}-\frac{p_ap_b}{2ab}\\+\frac{(n-2)}{4\omega a^2b^2}p_\phi^2\biggr].
\end{multline}
For fluid sector, Hamiltonian corresponding to Lagrangian (\ref{c2lf}) can be written as 
\begin{equation} \label{c2hf}
H_f=\frac{N}{a^{3\alpha}b^{(n-4)\alpha}}p_T
\end{equation}
where $T$ and $p_T$ are related by following canonical transformations
\begin{equation}
T=p_S e^{-S}p_\epsilon^{-(\alpha +1)},\hspace{0.5cm} p_T= p_\epsilon^{\alpha +1}e^S, \hspace{0.5cm}\bar{\epsilon}=\epsilon-(\alpha +1)\frac{p_s}{p_\epsilon},\hspace{0.5cm} \bar{p_\epsilon}=p_\epsilon.
\end{equation}
Using (\ref{c2hg}) and (\ref{c2hf}), the net Hamiltonian can be written as 
\begin{multline}\label{c2neth}
\mathcal{H} = \frac{N}{a^3b^{(n-4)}}\biggl[\frac{(n-5)a^2}{12(n-2)}p_a^2+\frac{b^2}{2(n-4)(n-2)}p_b^2 \\-\frac{ab}{2(n-2)}p_ap_b+\frac{1}{4\omega}p_\phi^2+a^{3-3\alpha}b^{(n-4)-(n-4)\alpha}p_T\biggr].
\end{multline}
 The above expression  for $\mathcal{H}$ may look formidable, but can be written in a simpler form with the help of following canonical transformations, 
 \begin{equation}
 A=lna, p_A=ap_a, B=lnb, p_B=bp_b,
\end{equation}
as
\begin{multline} \label{c2nth}
\mathcal{H} = \bar{N} \biggl[ \frac{(n-5)}{12(n-2)}p_A^2-\frac{1}{2(n-2)}p_A p_B+\frac{1}{2(n-2)(n-4)}p_B^2 \\ +\frac{1}{4\omega}p^2_\phi +e^{(3\alpha-3)A+((n-4)\alpha-(n-4))B}p_T \biggr],
\end{multline}
where $\bar{N}=\frac{N}{a^3b^{n-4}}$.
\section{Quantization of the model}
By varying the action with respect to $\bar{N}$, Hamiltonian constraint can be written as $\hat{H}=\frac{\mathcal{H}}{\bar{N}}=0$. Now the coordinates ($A, B, \phi$ and their corresponding conjugate momenta are promoted to operators using the standard commutation relations ($[q_i, p_i] = i$ in the units $\frac{h}{2\pi} = 1$). $\hat{H}$ can be written in terms of these operators using eq (\ref{c2nth}) and the Wheeler-DeWitt equation ($\hat{H}\psi=0$) looks like,
\begin{multline}\label{c2wd}
\biggl[ -\frac{(n-5)}{12(n-2)}\frac{\partial^2}{\partial A^2}  +\frac{1}{2(n-2)} \frac{\partial^2}{\partial A \partial B} -\frac{1}{2(n-2)(n-4)} \frac{\partial^2}{\partial B^2}  -\frac{1}{4\omega}\frac{\partial^2}{\partial \phi^2}\\ -ie^{(3\alpha-3)A+((n-4)\alpha-(n-4))B}\frac{\partial}{\partial T}\biggr]\psi(A,B,\phi, T)=0.
  \end{multline}
For $n=5$, the first term of equation (\ref{c2wd}) will vanish and Wheeler-DeWitt equation becomes simpler. We shall be considering it later. \\
With ansatz, $\psi(A,B,\phi, T)=\xi(A,B,\phi)e^{-iET}$,  equation (\ref{c2wd}) takes following form,
\begin{multline} \label{c2wdw}
\biggl[ \left(2-\sqrt{\frac{6(n-4)}{(n-5)}}\right)\frac{\partial^2}{\partial u^2}+\left(2+\sqrt{\frac{6(n-4)}{(n-5)}}\right)\frac{\partial^2}{\partial v^2}\\+\frac{1}{4\omega}\frac{\partial^2}{\partial \phi^2}+exp\biggl\{ (3\alpha-3)\biggl( \sqrt{\frac{n-5}{12(n-2)}}\frac{u+v}{2}\biggr)\\+((n-4)\alpha-(n-4))\biggl(  \sqrt{\frac{1}{2(n-2)(n-4)}}\frac{u-v}{2}\biggr)\biggr\}\\ E \biggr] \xi(u,v,\phi)=0,
\end{multline}
where $u$ and $v$ are given as,
\begin{eqnarray}
u=\sqrt{\frac{12(n-2)}{(n-5)}}A+\sqrt{2(n-2)(n-4)}B, \\
v=\sqrt{\frac{12(n-2)}{(n-5)}}A-\sqrt{2(n-2)(n-4)}B.
\end{eqnarray}
\subsection{For stiff  fluid $\alpha=1$ and $n>5$} 
For  $\alpha=1$ i.e. $P=\rho$, equation (\ref{c2wdw}) can be rewritten as
\begin{multline} \label{c2wdw1}
\biggl[ \left(2-\sqrt{\frac{6(n-4)}{(n-5)}}\right)\frac{\partial^2}{\partial u^2}+\left(2+\sqrt{\frac{6(n-4)}{(n-5)}}\right)\frac{\partial^2}{\partial v^2}\\+\frac{1}{4\omega}\frac{\partial^2}{\partial \phi^2}+E \biggr] \xi(u,v,\phi)=0.
\end{multline}
Solution for equation (\ref{c2wdw1}) can be written as 
\begin{multline}
\psi_{\lambda,E,k}(u,v,\phi,T)=K \sin(u\sqrt{\frac{\lambda}{|2-\bar{n}|}})\\ \times \sin(v\sqrt{\frac{E-k-\lambda}{2+\bar{n}}})  \sin(\phi\sqrt{4\omega k})e^{-i E T},
\end{multline} 
where $\bar{n}=\sqrt{6(n-4)/(n-5)}$ and $\lambda$, $k$ and $K$ are constant. Here we take $E>k+\lambda$.
By superposing the function $\psi_{\lambda,E,k}(u,v,\phi,T)$, the wave packet can be formed as;
\begin{multline}\label{c2wp}
\Psi(u,v,\phi,T)=\int_0^\infty d\lambda \int_0^\infty dk \int_0^\infty dE'\sin(u\sqrt{\frac{\lambda}{2-\bar{n}}})\\ \times\sin(v\sqrt{\frac{E'}{2+\bar{n}}})\sin(\phi\sqrt{4\omega k})e^{- (\lambda+E'+k) (\gamma+iT)},
\end{multline}
where $E'=E-k-\lambda$ and $\gamma$ is a positive constant.  We are choosing $e^{-\gamma (\lambda+E'+k) }$ as weight factor, which may not be unique for superposition, which gives us the normalized wave packet.  \\Norm $\| \Psi\|$ can be calculated as,
\begin{equation}
\| \Psi\|=\int \Psi \Psi^* d\bar{u}d\bar{v}d\phi=\frac{1}{32 \sqrt{2\omega}}\left(\frac{\pi}{\gamma}\right) ^{9/2},
\end{equation}
where $\bar{u}=\frac{u}{\sqrt{|2-\sqrt{6(n-4)/(n-5)}|}}$ and $\bar{v}=\frac{v}{\sqrt{2+\sqrt{6(n-4)/(n-5)}}}$.\\
This clearly shows that we have a finite and time-independent norm.
From equation (\ref{c2wp}), the normalized wave packet can be written as,
\begin{multline} \label{c2wvp}
\Psi(\bar{u},\bar{v},\phi,T)=\biggl( \frac{2\omega}{\pi} \biggr)^{3/4} \biggl( \frac{\sqrt{\gamma} }{\gamma+iT} \biggr)^{9/2}  \bar{u} \bar{v} \phi \\ \times exp \biggl[-\frac{1}{4}\biggl(\frac{4 \omega \phi^2+\bar{u}^2+\bar{v}^2}{ \gamma+i T}\biggr)\biggr].
\end{multline}
Now we can calculate the expectation values of $a$, $b$ and the proper volume measure $V= (-g)^{\frac{1}{2}}=a^3b^{n-4}$ in terms of $\bar{u}$ and $\bar{v}$ as,
\begin{multline}\label{c2ea}
\langle a \rangle =\int_{-\infty}^\infty d\bar{u}\int_{-\infty}^\infty d\bar{v} \int_{-\infty}^\infty d\phi |\Psi(\bar{u},\bar{v},\phi,T)|^2  exp\biggl[\frac{1}{2}\sqrt{\frac{n-5}{n-2}}\\ \times\biggl(\sqrt{\left|2-\sqrt{\frac{6(n-4)}{(n-5)}}\right|}\bar{u} +\sqrt{2+\sqrt{\frac{6(n-4)}{(n-5)}}}\bar{v}\biggr)\biggr] ,
\end{multline}
\begin{multline}\label{c2eb}
\langle b \rangle =\int_{-\infty}^\infty d\bar{u}\int_{-\infty}^\infty d\bar{v} \int_{-\infty}^\infty d\phi |\Psi(\bar{u},\bar{v},\phi,T)|^2  exp \biggl[\frac{1}{2}\sqrt{\frac{1}{(n-2)(n-4)}} \\ \times\biggl(\sqrt{\left|2-\sqrt{\frac{6(n-4)}{(n-5)}}\right|}\bar{u} -\sqrt{2+\sqrt{\frac{6(n-4)}{(n-5)}}}\bar{v}\biggr)\biggr],
\end{multline}
\begin{multline}\label{c2ev}
\langle V \rangle =\int_{-\infty}^\infty d\bar{u}\int_{-\infty}^\infty d\bar{v} \int_{-\infty}^\infty d\phi |\Psi(\bar{u},\bar{v},\phi,T)|^2 a^3(\bar{u},\bar{v})b^{n-4}(\bar{u},\bar{v}). 
\end{multline}
The plots for the expectation values are shown in the figures taking the value of positive constant $\gamma=1$.  
\begin{figure}[!t]
\centering
\includegraphics[scale=0.9]{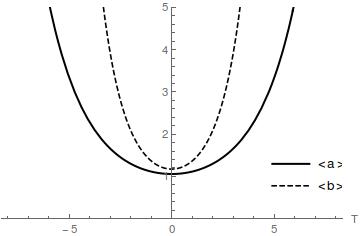}
\caption{Expectation value of scale factors for n=6.} \label{fig:1}
\end{figure}
\begin{figure}[!t]
\centering
\includegraphics[scale=0.9]{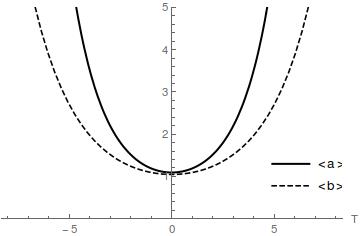}
\caption{Expectation value of scale factors for n=8 .} \label{fig:2}
\end{figure}
\begin{figure}[!t]
\centering
\includegraphics[scale=0.9]{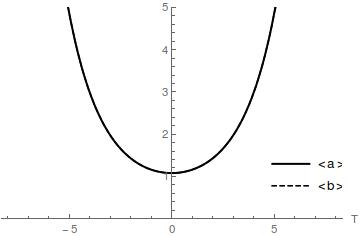}
\caption{Expectation value of scale factors for n=7 is given above and as it is shown, they coincide.} \label{fig:3}
\end{figure}
\begin{figure}[!t]
\centering
\includegraphics[scale=0.9]{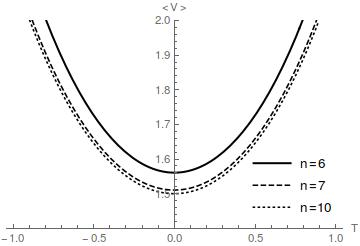}
\caption{Expectation value of Volume $<V=a^3b^{n-4}>$ .}\label{fig:4}
\end{figure}
The figures clearly indicate that there is no singularity of a zero proper volume. In fact both the scale factors $a$ and $b$ are quite well behaved individually.
\subsection{For stiff  fluid $\alpha=1$ and $n=5$}
For $n=5$ with $\alpha=1$,equation(\ref{c2wd}) will become 
\begin{multline} \label{c2wdw3}
\biggl[\frac{1}{6}\frac{\partial^2}{\partial A \partial B }-\frac{1}{6}\frac{\partial^2}{\partial B^2}-\frac{1}{4\omega}\frac{\partial^2}{\partial \phi^2}-i\frac{\partial }{\partial T} \biggr] \psi(A,B,\phi,T)=0.
\end{multline}
With the change of variable as $x=2A+B$ and $y=B$, it can be re-written as,
\begin{multline} \label{c2wdw2}
\biggl[-\frac{1}{6}\frac{\partial^2}{\partial x^2}+\frac{1}{6}\frac{\partial^2}{\partial y^2}+\frac{1}{4\omega}\frac{\partial^2}{\partial \phi^2}+ E \biggr] \xi(x,y,\phi)=0.
\end{multline}
Solution of above equation can be found using similar method what was done in the previous section. Normalized wave packet is given as,
\begin{multline}  \label{c2wp1}
\Psi(\bar{x},\bar{y},\phi,T)=\biggl( \frac{2\omega}{\pi} \biggr)^{3/4} \biggl( \frac{\gamma^{3/2} }{(\gamma^2+T^2)(\gamma+iT)} \biggr)^{3/2} \\ \times \bar{x} \bar{y} \phi exp \biggl[-\frac{1}{4}\biggl(\frac{\bar{x}^2}{ \gamma-i T}+\frac{4 \omega \phi^2+\bar{y}^2}{ \gamma+i T}\biggr)\biggr],
\end{multline}
where $\bar{x}=\sqrt{6}x$ and $\bar{y}=\sqrt{6}y$.\\
The relevant expectation values are given as 
\begin{eqnarray} \label{c2eab}
\langle a \rangle= \frac{e^{\frac{\gamma^2+T^2}{24 \gamma}} \left(\gamma (\gamma+24)+T^2\right)^2}{576 \gamma^2},\\\label{c2eabc}
\langle b \rangle=\frac{e^{\frac{\gamma^2+T^2}{12 \gamma}} \left(\gamma (\gamma+6)+T^2\right)}{6 \gamma}.
\end{eqnarray}
\begin{figure}[hbtp]
\centering
\includegraphics[scale=0.9]{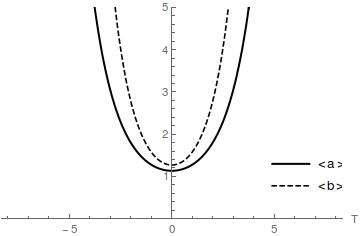}
\caption{Expectation value of scale factors for n=5.} \label{fig:5}
\end{figure}
\subsection{For fluid with $\alpha\neq 1$}
We will be considering  only $n=7$ as an example for a general fluid with $\alpha\neq 1$, in which case the equation is tractable. Now Wheeler-DeWitt equation (\ref{c2wdw}) can be written as,
\begin{multline} \label{c2wdw2}
\biggl[-\frac{\partial^2}{\partial u^2}+5\frac{\partial^2}{\partial v^2}+\frac{1}{4\omega}\frac{\partial^2}{\partial \phi^2}+e^{\frac{3\alpha-3}{\sqrt{30}}u} E \biggr] \xi(u,v,\phi)=0.
\end{multline}
With suitable operator ordering as  done in work by Pal and Banerjee\cite{spal1,spal2}, equation (\ref{c2wdw2}) takes following form, 
\begin{multline}
\biggl[ e^{\frac{(3-3\alpha)}{2\sqrt{30}}u}\frac{\partial}{ \partial u}e^{\frac{(3-3\alpha)}{2\sqrt{30}}u}\frac{\partial}{ \partial u}  -5e^{\frac{(3-3\alpha)}{\sqrt{30}}u}\frac{\partial^2}{\partial v^2} -e^{\frac{(3-3\alpha)}{\sqrt{30}}u} \frac{1}{4\omega}\frac{\partial^2}{\partial \phi^2}-E \biggl] \xi(u,v,\phi)=0.
\end{multline} 
Now with ansatz $\xi(u,v,\phi)=U(u)e^{iV_0v/\sqrt{5}}e^{i\sqrt{4\omega}\phi}$, above equation can be re-written as,
\begin{multline}
\biggl[ e^{\frac{(3-3\alpha)}{2\sqrt{30}}u}\frac{\partial}{ \partial u}e^{\frac{(3-3\alpha)}{2\sqrt{30}}u}\frac{\partial}{ \partial u} +e^{\frac{(3-3\alpha)}{\sqrt{30}}u}(V_0^2+k^2) -E \biggl] U(u)=0.
\end{multline}
With change of variable like $\chi= e^{\frac{(3\alpha-3)}{2\sqrt{30}}u}$, this equation can be written as,
\begin{equation} \label{c2henkel}
\biggl[ \frac{3(\alpha-1)^2}{40}\frac{\partial^2}{ \partial \chi^2}+\frac{V_0^2+k^2}{\chi^2} -E \biggr] U(\chi)=0 .
\end{equation}
Above equation is similar to the inverse square potential problem of physics which admits the self-adjoint extensions. The solution to equation(\ref{c2henkel}) can be given by Hankel functions. It has been extensively shown in the work\cite{spal1}.

\section{Discussion with Concluding Remarks}

In this work, using the n-dimensional generalization of Schutz formalism\cite{ndfrw}, we formulated the Lagrangian and Hamiltonian for our model. Followed by multiple canonical transformations, we obtained the complicated looking  Wheeler-DeWitt equation(\ref{c2wdw}) for the wave function of the n-dimensional anisotropic universe. We obtained the general wave packet of Wheeler-DeWitt equation for stiff fluid $\alpha=1$.  We are able to find a time independent and finite normed wave packet which establishes the unitary evolution of the model. 

We have found the non-singular expectation values of the scale factors and these are shown in Fig. \ref{fig:1}, \ref{fig:2}, \ref{fig:3} and \ref{fig:5} for different n.  Expectation values of scale factors $a$ and $b$ coincide for $n=7$.  This is not surprising as both usual space-section and the extra 3-dimensional space, although anisotropic between each other to start with, are 3-dimensional isotropic space like sections in themselves.
 
 Non-zero minima of the expectation values of volume element(Fig. \ref{fig:4}) clearly show the contraction of the universe followed by expansion avoiding any singularity indicating a bouncing universe. Similar results were obtained in quantization of anisotropic models in absence of fluid in the past \cite{ndaniso}.
  This result is obtained with quite reasonable boundary conditions, $\Psi \rightarrow 0$ for infinite values of its arguments, $\bar{u}$ and $\bar{v}$. For a discussion, we refer to the work of Vilenkin and Yamada \cite{nob} and Tuccii and Lehners \cite{nobb}. 

For a general fluid with $\alpha \neq 1$, we are able to show the self-adjoint extension and thus unitary evolution of the model for $n=7$ despite the computational difficulty of the model. However, the result of it is in accordance with four-dimensional anisotropic model Bianchi-I as expected \cite{spal1}.

Our results may also be considered as the higher dimensional generalization of unitary evolution of the anisotropic models as shown in \cite{spal1,spal2,spal3,sachin}.

\begin{titlepage}
\newpage
\clearpage
\newpage
\mbox{~}
\clearpage
\newpage
\end{titlepage}
\chapter[Unitarity in quantum cosmology: symmetries protected and violated]{Unitarity in quantum cosmology: symmetries protected and violated\footnote{ The work illustrated in this chapter is under consideration for publication; \cite{sachin4}.}}\label{ch4:1}

The Wheeler-DeWitt scheme of quantization\cite{wheeler1987superspace, dewitt1967quantum} of cosmological models was believed to have been plagued with the non-unitartity in anisotropic cosmological models\cite{pinto2013quantum, alvarenga2003}. In the first chapter, we have seen that there are ample examples \cite{spal1, spal2, spal3, spal4, almeida2017quantum, Majumder2013} of the possibility of a unitary evolution achieved by an operator ordering.. This unitarity is achieved by a suitable operator ordering. In fact, a theorem has recently been proved\cite{spaljmp} to show that at least for homogeneous models, it is always possible a have a self-adjoint extension for the Hamiltonian, and thus to have a unitary evolution for the system. The last two chapters, some different kinds of anisotropic models were shown to have similar features. \\

The purpose of this chapter is to look at the required price for the self-adjoint extension, in terms of symmetry. We shall look at two aspects, one is the Noether symmetry and the other being the  scale invariance. We shall deal with one example, the Bianchi I cosmological model, which is the simplest anisotropic cosmological model, but brings out the associated physical content quite comprehensively. \\

\section{Bianchi I space-time}

This example has already been worked out in an earlier work\cite{spal1}. We narrate the important steps here so as to make the presentation self sufficient.

We start with the action
\begin{equation}\label{c3action}
A = \int_M d^4x\sqrt{-g}R+2\int_{\partial M} d^4x\sqrt{h}h_{ab}K^{ab}+\int_M d^4x\sqrt{-g}P ,
\end{equation}
in a four dimensional space-time manifold M. R is the Ricci Scalar, $K^{ab}$ is extrinsic curvature and $h^{ab}$ is induced metric on the boundary $\partial M$. The first two terms correspond to the gravity sector and third term is due to a  perfect fluid which is taken as the matter constituent of the universe, $P$ is the pressure of the fluid. We have chosen our units, such that $16\pi G =1$. The second term will not contribute in the Euler-Lagrange equations as there is no variation in the boundary.\\

Bianchi I metric is given as
\begin{equation}\label{c3metric}
ds^2=n^2(t)dt^2-a^2(t)dx^2-b^2(t)dy^2-c^2(t)dz^2.
\end{equation}
With this metric, the gravity sector of the action can be written as 
\begin{equation}
\label{c3action-grav1}
\mathcal{A}_g=\int dt \bigg[-\frac{2}{n}[\dot{a}\dot{b}c+\dot{b}\dot{c}a+\dot{c}\dot{a}b]\bigg].
\end{equation}

A transformation of variables as
\begin{eqnarray}
\label{c3trans-1}
a(t)=e^{\beta_0+\beta_++\sqrt{3}\beta_-}, \\
b(t)=e^{\beta_0+\beta_+-\sqrt{3}\beta_-}, \\
c(t)=e^{\beta_0-2\beta_+},
\end{eqnarray}
will make the Lagrangian in equation (\ref{c3action-grav1}) look like,
\begin{equation}
\label{c3lagrangian}
\mathcal{L}_g = -6\frac{e^{3\beta_0}}{n}[\dot{\beta}_0^2-\dot{\beta}_+^2-\dot{\beta}_-^2].
\end{equation}

The canonically conjugate momenta are defined as usual as $p_i = \frac{\partial {\mathcal{L}_g}}{\partial x_i}$. The Hamiltonian for the gravity sector looks like
\begin{equation}\label{c3hamgrav}
 H_{g}=-n\exp(-3\beta_{0})\left\{\frac{1}{24}\left(p_{0}^{2}-p_{+}^{2}-p_{-}^{2}\right)\right\}.
\end{equation}

For the fluid sector, the practice is to adopt the Schutz formalism of expressing the fluid properties like density and pressure in terms of some thermodynamic variables and then effect a set of canonical transformations. The method is described elaborately  in the introduction. The relevant action is
\begin{equation}
\label{c3action-fluid}
\begin{split}
\mathcal{A}_{f}&=\int dt \mathcal{L}_{f}\\&= \int dt \left[n^{-\frac{1}{\alpha}}e^{3\beta_{0}}\frac{\alpha}{\left(1+\alpha\right)^{1+\frac{1}{\alpha}}}\left(\dot{\epsilon}+\theta\dot{S}\right)^{1+\frac{1}{\alpha}}e^{-\frac{S}{\alpha}}\right],
\end{split}
\end{equation}
 where $\epsilon, \theta$ and $S$ are the thermodynamic quantities (Schutz variables), and $\alpha$ is a constant that connects the density ($\rho$) and pressure ($P$) as $P = \alpha \rho$. \\

As the metric components do not depend on spatial coordinates, the spatial volume integrates out as a constant and will not participate in the subsequent calculations. Also, the boundary term is ignored as that does not contribute to the variation of the action.

The Canonical momenta are defined as $p_{\epsilon}=\frac{\partial\mathcal{L}_{f}}{\partial\dot{\epsilon}}$ and
$p_{S}=\frac{\partial\mathcal{L}_{f}}{\partial\dot{S}}$ and execute a set of canonical transformations,
\begin{eqnarray}\label{c3can-trans}
T&=&-p_{S}\exp(-S)p_{\epsilon}^{-\alpha -1},\\
p_{T}&=&p_{\epsilon}^{\alpha+1}\exp(S),\\
\epsilon^{\prime}&=&\epsilon+\left(\alpha+1\right)\frac{p_{S}}{p_{\epsilon}},\\
p_{\epsilon}^{\prime}&=&p_{\epsilon}.
\end{eqnarray}

The Hamiltonian for the fluid becomes 
\begin{equation}
\label{c3ham-fl}
H_{f}= ne^{-3\beta_{0}}e^{3\left(1-\alpha\right)\beta_{0}}p_{T}.
\end{equation}

One can now write the net Hamiltonian as 
\begin{equation}\label{c3ham-net}
H=n\exp(-3\beta_{0})\left\{-\frac{1}{24}\left(p_{0}^{2}-p_{+}^{2}-p_{-}^{2}\right)+e^{3(1-\alpha)\beta_{0}}p_{T}\right\},
\end{equation}
and proceed to quantize the system by using the Hamiltonian constraint $H=0$ (obtained by varying the action with respect to the lapse function $n$) and raising the variables as operators, which yields the Wheeler-DeWitt equation as 
\begin{equation}\label{c3WDeq}
\left(\frac{\partial^{2}}{\partial \beta^{2}_{0}}-\frac{\partial^{2}}{\partial\beta^{2}_{+}}-\frac{\partial^{2}}{\partial\beta^{2}_{-}}\right)\psi = 24\imath e^{3(1-\alpha)\beta_{0}}\frac{\partial \psi}{\partial T}.
\end{equation} 

In writing this equation, a choice of gauge has been made ($n=e^{3\alpha{\beta}_{0}}$). With the usual separability ansatz $\psi(\beta_{0},\beta_{+},\beta_{-},T)=\phi(\beta_{0},\beta_{+},\beta_{-})e^{-\imath ET}$ where $E$ is a constant, the equation (\ref{c3WDeq}) becomes 
\begin{equation}\label{c3WDeq2}
\left(\frac{\partial^{2}}{\partial \beta^{2}_{0}}-\frac{\partial^{2}}{\partial\beta^{2}_{+}}-\frac{\partial^{2}}{\partial\beta^{2}_{-}}\right)\phi =24E\phi e^{3(1-\alpha)\beta_{0}}.
\end{equation}

For the general case $0\leq \alpha \leq 1$, choose another separation of variables as $\phi = \xi (\beta_0) \eta (\beta_{+} , \beta_{-})$ and a coordinate transformation as $\chi = e^{-\frac{3}{2}(\alpha-1)\beta_{0}}$, a long but straightforward calculation will yield an equation for $\xi$ as 
\begin{equation}\label{c3xi}
-\frac{d^{2}\xi}{d\chi^{2}}-\frac{\sigma}{\chi^{2}}\xi = -E^{\prime}\xi,
\end{equation} 
where $E^{\prime}=\frac{32}{3\left(1-\alpha \right)^{2}}E$ and $\sigma$ is composed of the separation constants coming in the process of separation of the functions of $\beta_{+}$ and $\beta_{-}$.  This equation indeed has a favourable deficiency index, which guarantees  the existence of a self-adjoint extension\cite{vaughn} and thus, the evolution of the system is unitary. 

The work of Pal and Banerjee\cite{spal1} shows that this transformation, even at the classical level, gives rise to a Hamiltonian which, when raised to operators, is self-adjoint.  It deserves mention that often the unitarity is achieved by means of an operator ordering, which is not unique\cite{spaljmp}. Thus the most unambiguous example would be the one if one can effect a coordinate transformation at the classical level so that the operator ordering is irrelevant at the quantum level. The present example is exactly that and this is one good reason for choosing Bianchi I at the outset. For a very brief review of various aspects of factor ordering, we refer to \cite{Tavakoli_2019}. 

\section{Preservation of Noether symmetry}

The generator for Lagrangian (\ref{c3lagrangian}) can be written as: 
 \begin{equation}
 \textbf{X}=b_0\frac{\partial}{\partial\beta_0}+b_+\frac{\partial}{\partial\beta_+}+b_-\frac{\partial}{\partial\beta_-}+\dot{b_0}\frac{\partial}{\partial\dot{\beta_0}}+\dot{b_+}\frac{\partial}{\partial\dot{\beta_+}}+\dot{b_-}\frac{\partial}{\partial\dot{\beta_-}},
 \end{equation}
where $b_i (\beta_j)$s ($i, j = 0, +, -$) are to be determined from the Noether symmetry condition
\begin{equation}
\label{c3noether}
 \pounds_\textbf{X}\mathcal{L}_g=0,
\end{equation}
meaning the Lie derivative of the Lagrangian with respect to $\textbf X$ is zero. \\
This condition yields the set of equations
\beq
 \frac{3}{2}b_0+\frac{\partial b_0}{\partial \beta_0}=0,\eeq\beq
  \frac{3}{2}b_0+\frac{\partial b_+}{\partial \beta_+}=0,\eeq\beq
   \frac{3}{2}b_0+\frac{\partial b_-}{\partial \beta_-}=0.
 \eeq

The solution for this set of equations can be written as 
\begin{eqnarray} \label{c3con1}
 e^{\frac{3}{2}\beta_0}b_0=Q_1=constant, \\
\frac{3}{2}\beta_+b_0+b_+=Q_2=constant, \\
\frac{3}{2}\beta_-b_0+b_-=Q_3=constant.
\end{eqnarray}

 A coordinate transformation of the form $\chi=e^{-\frac{3}{2}(\alpha-1)\beta_0}$ transforms the Lagrangian density given in (\ref{c3lagrangian}) to the form

\begin{equation}\label{c3ltran}
\mathcal{L}_{g_T}=-\frac{6}{n}[\frac{4\chi^{2\alpha/(1-\alpha)}}{9(1-\alpha)^2}\dot{\chi^2}-\chi^{2/(1-\alpha)}(\dot{\beta}_+^2+\dot{\beta}_-^2)]. 
\end{equation}

It is now required to check whether the Noether symmetry corresponding to $\mathcal{L}_g$ is retained in $\mathcal{L}_{g_T}.$

The corresponding generator for the Lagrangian as in equation (\ref{c3ltran}) can be written as,
\begin{equation}
\textbf{X}=q_0\frac{\partial}{\partial\chi}+q_+\frac{\partial}{\partial\beta_+}+q_-\frac{\partial}{\partial\beta_-}+\dot{q_0}\frac{\partial}{\partial\dot{\chi}}+\dot{q_+}\frac{\partial}{\partial\dot{\beta}_+}+\dot{q}_-\frac{\partial}{\partial\dot{\beta}_-},
\end{equation}
where $q_i (\chi,\beta_j)$ ($i,j = 0,+,-$) are to be determined from the Noether symmetry condition $\pounds_\textbf{X}\mathcal{L}_{g_T}=0$, which in this case gives following partial differential equations,
 \begin{eqnarray}
 \frac{\alpha}{1-\alpha}\frac{q_0}{\chi}+\frac{\partial q_0}{\partial \chi}=0,\\
 \frac{\alpha}{1-\alpha}\frac{q_0}{\chi}+\frac{\partial q_+}{\partial \beta_+}=0,\\
 \frac{\alpha}{1-\alpha}\frac{q_0}{\chi}+\frac{\partial q_-}{\partial \beta_-}=0.
 \end{eqnarray}

 Solution to above three equations can be given as,
 \begin{eqnarray} \label{c3con1}
 {q_0}^{1/\alpha}\chi^{1/(1-\alpha)}=Q_A=constant, \\
\frac{q_0 \beta_+}{(1-\alpha)\chi}+q_+=Q_B=constant, \\
\frac{q_0\beta_-}{(1-\alpha)\chi}+q_-=Q_C=constant.
 \end{eqnarray}

It is easy to check that the  solution to Noether symmetry conditions for both Lagrangian match exactly with the identification $b_0=q_{0}^{1/\alpha}=[\frac{3}{2}(1-\alpha)]^{1/(\alpha-1)}e^{-\frac{3}{2}\beta_0}.$ \\

It should be noted that a self-adjoint extension actually involves working on the Hilbert space\cite{vaughn}. But it has been brought about by ordering of operators\cite{spal1}. As there is indication of the uniqueness of the operator ordering to achieve this\cite{spaljmp}, a safe way is to get a coordinate at the classical level, such that while promoting the variables to operators for the quantization, the ordering is at least unambiguous. The coordinate $\chi$ exactly serves this purpose.
\section{Loss of scale invariance}
We shall now look at the issue of scale invariance. Under a scale transformation,  there is a rule of transformation of the coordinate and time, e.g., if $x=x(t)$ is transformed like $\bar{x} = \lambda^{-1/2} x$, $t$ should go like $\bar{t} =\lambda t$, where $\lambda$ is the scale. 

We now look back at the equation (\ref{c3xi}), where we easily identify $\chi$ and $T$ as the coordinate and time respectively. So the relevant transformations will be $\bar{T}=\lambda T$, $\bar{\chi}(\bar{T})=\lambda^{-1/2}\chi(\lambda T)$. Energy  will transform $\bar{E}'=E'/\lambda$ and $\frac{\partial}{\partial{\bar{\chi}}}=\lambda^{1/2}\frac{\partial}{\partial{\chi}}$. For a comprehensive description of this, we refer to references  \cite{cabo} and \cite{pradhan}.

If we effect this scale transformation, the equation (\ref{c3xi})takes the form as
 \begin{equation}
 -\lambda\frac{d^2\xi}{d{\chi}^2}-\lambda\frac{\sigma}{{\chi}^2}\xi-=-\frac{E'}{\lambda}\xi.
 \end{equation}

This clearly does not preserve the scale invariance! However, this should not perhaps be considered too costly, as there is indeed an incompatibility between hermiticity and scale invariance in general. This has been proved very elegantly by Pal\cite{spal-scale}. So this is not at all an artefact of anisotropic quantum cosmology.

\section{Conclusion}

It is now known that the allaged non-unitarity of the anisotropic quantum cosmological models is not true. Anistropic models, at least if they are spatially homogeneous, are shown to have self-adjoint extension\cite{spaljmp}. The present work looks at the cost of this extension in terms of symmetry. \\

With the example of the Bianchi I metric, it is quite clearly shown that the self-adjoint extension indeed retains the Noether symmetry. However, the scale invariance is clearly lost. So apparently, this is the cost  of a self-adjoint extension. As already mentioned, this does not appear to be too much of a price for securing unitarity as this feature is quite generic\cite{spal-scale}. \\
\chapter[Jordan and Einstein frame: Are they same at quantum level?]{Jordan and Einstein frame: Are they same at quantum level? \footnote{ The work illustrated in this chapter is published; \textbf{S. Pandey} and N. Banerjee , European Physical Journal Plus, \textbf{132}, 107(2017).}} \label{ch5:1}

A non-minimally coupled theory of gravity, where a  field interferes with the curvature scalar, has two popular frameworks for its description. One is called the Jordan frame, where the theory is manifestly non-minimal in the sense that the interference term is visible in the action and also in the field equations derived from the action by means of a variational principle. In the second framework, known as the Einstein frame, the non-minimal coupling is broken by means of a conformal transformation of the form $\bar{g}_{\mu\nu}=\Omega^2 g_{\mu\nu}$, the theory appears to be simpler and looks similar to General Relativity where the non-minimally coupled field appears as an additional term in the matter sector. In Jordan frame, the Newtonian constant of gravity G becomes a variable. Einstein's frame has a restored constancy of G but the rest mass of the test particle becomes a function and thus, one has to pay a bigger price, the validity of equivalence principle, and hence the significance of the geodesic equation is lost. In fact, this loss of the principle of equivalence is the key to understand the nature of the non-minimal coupling in spite of the apparent resemblance with general relativity.\\
The question of equivalence of these two frames for the same theory of gravity is yet to be settled. The apparent mismatch of the two are quite obvious; Jordan frame description rests heavily on the principle of equivalence, whereas the other does not respect that principle. The usual debate is centred around the question which frame is more dependable for the description of gravity. Cho indicated that the Einstein frame is more trustworthy for a physical description of gravity \cite{cho1992}.\\
However, Faraoni and Gunzig\cite{faraoni1999} indicated that in the classical regime, Jordan frame is the reliable one on the consideration of gravitational waves. Chiba and Yamaguchi\cite{chiba} estimated various cosmological parameters in these frames and showed that they are different and hence indeed a matter of concern.\\
Some investigations, however, show that the apparent discrepancy in the results obtained in the two frames can actually be resolved. Salgado resolved the mismatch between the Cauchy problem in the two frames \cite{salgado}. Faraoni and Nadeau argued that the nonequivalence of the two frames actually comes down to a matter of interpretation, at least at the classical level\cite{faraoni2007}. In the context of Higgs' inflation, Postma and Volponi showed an equivalence of Jordan and Einstein frames\cite{postma}.\\
Artymowski, Ma and Zhang\cite{ma} showed that Brans-Dicke theory looks different in the two frameworks in the context of loop quantum cosmology both in the presence or absence of another scalar field as the matter sector.\\
The question of equivalence is properly posed in the following way. The solutions obtained in one frame should be transformed into the second frame by means of the conformal transformation through which the metric components in the two versions are related and should then be compared with the solutions in the second frame. This would indicate whether the character of the frame itself introduces any feature which is not there in the other.\\
In this chapter, we ask this question of equivalence at the quantum level. We work with Brans-Dicke theory\cite{brans}, easily the most talked about theory amongst the nonminimally coupled theories of gravity. In this theory, a scalar  field $\phi$ is coupled with the Ricci scalar R in the action. We work in a spatially  flat, homogeneous and isotropic cosmological model in vacuum, and quantize the model following the standard canonical Wheeler-deWitt quantization scheme\cite{dewitt1967quantum,wheeler1987superspace}, and form the relevant wave packet  $\Psi_{Jordan}$ in Jordan frame. The action is then written in the Einstein frame via the conformal transformation $\bar{g}_{\mu\nu}=\phi g_{\mu\nu}$ suggested by Dicke\cite{dicke}. We pretend that this is a completely different theory and quantize a same cosmological model following the same Wheeler-DeWitt scheme. The wave packet  $\Psi_{Einstein}$ is formed. Naturally, it looks different from the wave packet in the Jordan frame. We now effect the inverse transformation in $\bar{g}_{\mu\nu}$ in the wave packet  $\Psi_{Einstein}$, and see that it is exactly the same as  $\Psi_{Jordan}$. The result is quite general,in the sense that this does not depend upon the parameter of the theory $\omega$.\\
In the next two sections, the quantization of the cosmological model in vacuum is discussed in Jordan and Einstein frames respectively. Finally, the result is critically analyzed in the section \ref{sec:l}.

\section{Jordan frame} 
The relevant action in Brans Dicke theory without any contribution from the matter sector, in the so-called Jordan frame, is written as
\begin{equation} \label{c4act}
A=\int d^4 x \sqrt{-g} \bigg{[} \phi R + \frac{\omega}{\phi}\partial_\mu \phi \partial^\nu \phi \bigg{]} ,
\end{equation}
where R is the Ricci scalar, $\phi$ is the scalar  field and $\omega$ is a dimensionless parameter. It is generally believed that the higher the value of $\omega$, the closer the theory is to general relativity, and for $\omega \to \infty $, the two theories (GR and BD) are identical. However, it has been proved that this equivalence of the two theories is not at all generic\cite{nbsen,far1999}.

A spatially homogeneous and isotropic space-time with a  flat spatial section is given as
\begin{equation}
ds^2 = n^2(t)dt^2 - a^2(t)dl^2,
\end{equation}
where the lapse function $n$ and the scale factor $a$ are functions of the time alone.  With this metric, the Lagrangian can be extracted from the action (\ref{c4act}) as
\begin{equation}
L=- \frac{6 \phi a \dot{a}^2}{n} -\frac{6 \dot{a} \dot{\phi}a^2}{n} + \frac{\omega}{n\phi}\dot{\phi}^2a^3.
\end{equation}
With the change of variable as 
\beq
a(t)=e^{-\alpha/2 +\beta},\eeq \beq
\phi(t) = e^{\alpha}
\eeq
Lagrangian can be written as
\begin{equation}
L = \frac{e^{-\alpha / 2 + 3 \beta}}{n}\bigg{[}-6\dot{\beta}^2+\frac{2\omega +3}{2}\dot{\alpha}^2 \bigg{]}.
\end{equation}
The corresponding Hamiltonian comes out to be 
\begin{equation}
H =  ne^{\alpha / 2 - 3 \beta}\bigg{[}-p_\beta^2+\frac{12}{2\omega+3}p_\alpha^2\bigg{]}.
\end{equation}
By a variation of the action in the first order with respect to the lapse function $n$, one has the Hamiltonian
constraint as $\mathcal{H}=\frac{H}{e^{-\alpha / 2 + 3 \beta}}=0$.
Now we consider a canonical transformation of variables as  $(\alpha,p_\alpha)$ to $(T,p_T)$ given by
\beq
T=\frac{\alpha}{p_\alpha},\eeq \beq
p_T=\frac{p_\alpha^2}{2}.
\eeq
It is easily verified that T and p T are canonically conjugate variables.\\
 One can now write  $\mathcal{H}$ as 
\begin{equation}
\mathcal{H}=-p_\beta^2+\frac{24}{2\omega+3}p_T.
\end{equation}
Here $\beta$, T are the coordinates and $p_\beta$ ,$p_T$ are the corresponding canonically conjugate momenta. The canonical structure can be verified from the relevant Poisson brackets.\\
The Wheeler-DeWitt (WDW) equation, $\mathcal{H}\psi = 0$, can be written as
\begin{equation} \label{c4wdw1}
\bigg{[}\frac{\partial^2}{\partial \beta^2} -i\frac{24}{2\omega+3} \frac{\partial}{\partial T}\bigg{]}\psi=0.
\end{equation}
Then solution for above equation is obtained as
\begin{equation}
\psi_E(\beta, T)= e^{iET}\sin[\sqrt{24E/(2\omega+3)}\beta],
\end{equation}
or
\begin{equation}
\psi_E(a,\phi, T)= e^{iET}\sin[\sqrt{24E/(2\omega+3)}ln(\sqrt{\phi}a)],
\end{equation}
where E is a constant of separation.\\
Using $\int_0^{\infty}e^{-\gamma x}\sin{\sqrt{mx}}dx=\frac{\sqrt{\pi m}}{2 \gamma^{3/2}}e^{-m/4\gamma}$, wave packet  can be written as
\begin{equation}\label{c4wpj}
\Psi(a,\phi,T)= \sqrt{\frac{6\pi}{2\omega+3}}\frac{ln(\sqrt{\phi}a)}{(\gamma-iT)^{3/2}}exp\bigg{[}-\frac{6 ln^2(\sqrt{\phi}a)}{(2\omega+3)(\gamma-iT)}\bigg{]}.
\end{equation}

\section{Einstein frame}
If one effects a conformal transformation given by
\begin{equation}
\bar{g}_{\mu\nu}=\Omega^2 g_{\mu\nu},
\end{equation}
the action will look like
\begin{equation}
A=\int d^4x \sqrt{-\bar{g}}\bigg{[}\bar{R}+\frac{2\omega+3}{2}\partial_\mu \xi \partial^\nu \xi \bigg{]},
\end{equation}
where $\xi = \ln\phi$\cite{dicke}. 
The Lagrangian in this case can be written as 
\begin{equation}
L=-\frac{6\dot{\bar{a}}^2\bar{a}}{\bar{n}}+\frac{2\omega+3}{2\bar{n}}\dot{\xi}^2\bar{a}^3,
\end{equation}
and the corresponding Hamiltonian becomes
\begin{equation}
H =(\bar{n}/\bar{a}^3) \bigg{[}-(\bar{a}^2/24)p_{\bar{a}}^2+\frac{1}{2(2\omega+3)}p_\xi^2\bigg{]}.
\end{equation}
The Hamiltonian constraint, as usual, can be obtained by varying the action with respect to the lapse function $\bar{n}$ as, $\mathcal{H}=\bar{a}^3H =0$.\\
 Again with a similar canonical transformation as
\beq
\bar{T}=\frac{\xi}{p_\xi},\eeq\beq
p_{\bar{T}}=\frac{p_\xi^2}{2},
\eeq
then WDW equation can be written as
\begin{equation}\label{c4wd}
\bigg{[}\bar{a}^2\frac{\partial^2}{\partial \bar{a}^2} -i\frac{24}{2\omega+3} \frac{\partial}{\partial \bar{T}}\bigg{]}\bar{\psi}=0.
\end{equation}
One can easily see from the transformation equations that the scalar time parameters in the two frames,$T$ and $\bar{T}$ are actually equal.
With an operator ordering of the first term on the left-hand side $\bar{a}\frac{\partial}{\partial \bar{a}}\bar{a}\frac{\partial}{\partial \bar{a}}$ and taking $\chi=\ln\bar{a}$, eq. (\ref{c4wd}) can be written as
\begin{equation}\label{c4wd}
\bigg{[}\frac{\partial^2}{\partial \chi} -i\frac{24}{2\omega+3} \frac{\partial}{\partial \bar{T}}\bigg{]}\bar{\psi}=0.
\end{equation}
Then solution for above equation can be given as
\begin{equation}
\bar{\psi}_E(\bar{a},\bar{T})=e^{iE\bar{T}}sin[\sqrt{\frac{24E}{(2\omega+3)}}ln(\bar{a})].
\end{equation}
The corresponding wave packet is
\begin{equation} \label{c4wpe}
\Psi(\bar{a},\bar{T})= \sqrt{\frac{6\pi}{2\omega+3}}\frac{ln(\bar{a})}{(\gamma-i\bar{T})^{3/2}}exp\bigg{[}-\frac{6ln^2(\bar{a})}{(2\omega+3)(\gamma-i\bar{T})}\bigg{]}.
\end{equation}
If one now reverts the conformal transformation and go back to the Jordan frame, by using $\bar{a}^2=a^2\phi$ and $\xi= ln \phi$, it is quite easy to see that the wave packet given in equation (\ref{c4wpe}) in the Einstein frame is exactly same as that in the Jordan frame given in equation (\ref{c4wpj}).

The criticism of this proof could well be that in order to separate the variables in the Jordan frame, a transformation has been used which is connected to the conformal transformation that takes the action to the Einstein frame! Thus the calculations are actually in the same Einstein frame. But a closer scrutiny will reveal that the transformation, being a point transformation, is a canonical transformation, so the equivalence is actually built in.

\section{Discussion} \label{sec:l}
The result obtained carries a clear message. If the action is not contaminated with other  fields, such as a fluid, the Jordan and Einstein frames are completely equivalent in the sense that one can go from one description to the other at the  final stage, i.e., at the level of the solution via the conformal transformation. This result is completely independent of the choice of the coupling constant $\omega$, which actually determines the deviation of the theory from general relativity. The work is carried out in Brans-Dicke theory. Of course, there are other more complicated non-minimally coupled theories where this has to be verified, but the message is significant.\\
Very recently a result contrary to this has been given\cite{nb}, where it was shown that the wave packets in the two frames behave in different ways even after  $\Psi_{einstein}$ is transformed back to the Jordan frame. The solutions were obtained for particular values of the Brans-Dicke parameter $\omega$, but that should not infringe upon the result. Perhaps the addition of a contribution of a fluid in the action results in the requirement of an ordering of operators, as the fluid variable and the geometry cannot be separated efficiently.\\

Furthermore, the conformal transformation, $\bar{g}_{\mu\nu}=\Omega^2 g_{\mu\nu}$, inflicts a change of units in the variables as indicated by Dicke\cite{dicke}, so one has to be careful about the interpretation of the results as shown by Faraoni and Nadeau\cite{faraoni2007}. For various choices of units and their significance, we also refer to the early work by Morganstern\cite{morgan}.\\

It deserves mention that as the cosmic time t is a coordinate and not a scalar parameter, the evolution of the quantum system requires a properly oriented scalar time parameter in the scheme, which is very efficiently constructed out of the fluid parameters as shown by Lapchinski and Rubakov\cite{Lapchinskii}. In the present work, as no fluid is considered, the scalar time parameter ($T$ and $\bar{T}$ respectively in the two frames) is constructed from the scalar field and the scale factor following the work of Vakili\cite{vakili2012scalar}. The derivatives with respect to $T$ and $\bar{T}$ appear in the first order in the Hamiltonian (equations (\ref{c4wdw1}) and (\ref{c4wd}) ), indicating their role in the scheme as time. One can easily check that the Poisson brackets $ \{ T, H\}$ and $ \{ \bar{T}, \bar{H} \} $have the correct signatures which ensure the proper orientation of the time parameter. For a detailed description of this issue, we refer to the recent work of Pal and Banerjee\cite{spal1}.\\

The present example discussed in this chapter  is definitely a particular theory, namely the Brans-Dicke theory. But the calculations are clean and the results are so unambiguous and general (independent of the Brans-Dicke coupling parameter $\omega$), that one can claim with confidence that the equivalence of the two frames are established, at the quantum level, at least when the action is taken in the pure form, i.e., without any matter field. 

\chapter[Equivalence of Jordan and Einstein frames in anisotropic quantum cosmological models]{Equivalence of Jordan and Einstein frames in anisotropic quantum cosmological models \footnote{ The work illustrated in this chapter is published; \textbf{S. Pandey}, S. Pal and N. Banerjee, Annals of Physics, \textbf{393}, 93(2018).} } \label{ch6:1}

In the previous chapter, we established the mathematical equivalence of Jordan and Einstein frames at a quantum level using an isotropic cosmological model. In this chapter, we shall discuss this equivalence using more involved anisotropic models.

Although the physical inequivalence and henceforth the question of which frame is the physical one is a moot point for decades\cite{morgan, magnano, sokol, cho1994, ippo, FGN}, mathematical equivalence of these two frames at a classical level has almost been taken for granted. By mathematical equivalence, we mean the following: we evaluate a quantity $T_{J}$ in the Jordan frame, evaluate the same quantity in the Einstein frame and obtain $T_{E}$,  mathematical equivalence is defined to be the statement that $T_{E}$ is just a transformed version of $T_{J}$. Given this scenario, it is meaningful to explore whether this mathematical equivalence survives a quantization process. \\

In the light of recent resurgence in the Wheeler-DeWitt (WDW) quantization \cite{spal1,spal2,spal3,spal4,spaljmp,sachin, alvarenga2017} scheme, we have a natural framework to answer the question of equivalence at the quantum level. In fact, within the WDW quantization scheme \footnote{Similar non-equivalence is shown in the framework of loop quantum cosmology as well\cite{ma}}, there has been claim of non-equivalence at quantum level in literature \cite{nb}, albeit for cosmological models with a matter content. Very recently Kamenshchik and Steinwachs\cite{kamen}, with an estimate of the one loop divergence in the two frames, showed that the frames are not equivalent. Nonetheless, the debate seems to be open enough. The hint of quantum equivalence is also speculated in \cite{spal5} and further claimed to be true for isotropic homogeneous model with scale invariant matter content in \cite{almeida2017quantum} and quite generically in \cite{ndaniso}. The purpose of this work is to settle the issue and show that a consistent operator ordering can be chosen in the two frames so that the frames become equivalent. This ordering has to do with the ordering of generalized position and conjugate momenta in the Hamiltonian, once we promote the classical Hamiltonian to a quantum one and position, momenta become non-commuting operators. For every choice of an ordering in one frame, there is a particular choice of ordering in another frame, such that quantum Hamiltonian written in terms of operators become equivalent. In particular, we explicitly choose a consistent set of parametrization and operator ordering to show the equivalence for homogeneous models without matter content. 

The rest of the chapter is organized as follows. In section \ref{sec:1}, we analyze the model with zero spatial curvature, namely a Bianchi I model in detail. The techniques and method of the proof to study Bianchi-I is further generalized in the concluding section. We conclude our work with a general discussion along with a sketch of a general proof of equivalence between the two frames for all homogeneous models. It deserves mention that our result agrees with the work in references \cite{ndaniso} and \cite{barreto}. Models which are anisotropic generalization of constant non-zero spatial curvature isotropic models, i.e., Bianchi-V and Bianchi-IX are discussed in section \ref{sec:2}. Section \ref{sec:3} contains models with no isotropic analogue and  discussion on Locally rotationally symmetric Bianchi-I and Kantowski-Sachs models, respectively. The following table \ref{tab:spcu} shows the curvature of the spatial slice,  for the various models discussed in this chapter.
\begin{table}
\begin{center}
 \begin{tabular}{|c|c|} 
 \hline
 \textbf{Anisotropic Models} &  \textbf{ Spatial Curvature}  \\  
 \hline
 Bianchi-I & $ 0$ \\ 
 \hline
 Bianchi-V & $\frac{6m^2}{a^2(t)}$ \\
 \hline
 Bianchi-VI & $\frac{6(m^2-m+1)}{a^2(t)}$ \\
 \hline
 Bianchi-I(LRS) & $\frac{6}{h^2 a^2(t)}$ \\
 \hline
 Kantowski-Sachs & $-\frac{2}{b^2(t)}$ \\  
 \hline
\end{tabular}
\end{center}
\caption{Spatial curvature of various models}
\label{tab:spcu}
\end{table}

\section{Bianchi I} \label{sec:1}
In the absence of any matter field, action for Brans Dicke theory in the Jordan frame is given by
\begin{equation}\label{actionb1}
A_{J}=\int d^{4}x\mathcal{L}=\int d^4 x \sqrt{-g} \bigg{[} \phi R + \frac{\omega}{\phi}\partial_\mu \phi \partial^\nu \phi \bigg{]},
\end{equation}

where $\omega$ is the dimensionless Brans-Dicke coupling parameter and $\phi$ is the scalar field which depends only on time $t$.
The Bianchi-I metric is described by 
\begin{equation}
\label{b1metric}
ds^2 = n^2dt^2 - a^2(t)dx^2-b^2(t)dy^2-c^2(t)dz^2,
\end{equation}
where $a,b,c$ are the scale factors along three spatial direction. They encode the anisotropy present in the metric. 

The Lagrangian arising out of Eq.~(\ref{actionb1}) can be explicitly written as: 
\begin{equation}
L_{J}=-\frac{2\phi abc}{n}\left[ \frac{\dot{a}}{a} \frac{\dot{b}}{b}+ \frac{\dot{a}}{a} \frac{\dot{c}}{c}+\frac{\dot{b}}{b} \frac{\dot{c}}{c}+\frac{\dot{a}}{a}\frac{\dot{\phi}}{\phi}+ \frac{\dot{b}}{b} \frac{\dot{\phi}}{\phi}+\frac{ \dot{c}}{c} \frac{\dot{\phi}}{\phi}  - \frac{\omega}{2}\frac{\dot{\phi}^2}{\phi^{2}}\right],
\end{equation}
where an overhead dot signifies a differentiation with respect to $t$. We reparametrize the scale factors as,
\begin{eqnarray}
\label{b1trans}
a(t)&=&e^{\sigma_0 +\sigma_+ +\sqrt{3}\sigma_-},\\
b(t)&=&e^{\sigma_0 +\sigma_+ -\sqrt{3}\sigma_-},\\
c(t)&=&e^{\sigma_0 -2\sigma_+},\\
\phi(t)&=&e^{\alpha},
\end{eqnarray}
which recast the Lagrangian in following form:
\begin{equation}\label{L1}
L_{J}=\frac{e^{\alpha+3\sigma_{0}}}{n}\left[-6(\dot{\sigma}_{0}^{2}-\dot{\sigma}_{+}^{2}-\dot{\sigma}_{-}^{2})-6\dot{\alpha}\dot{\sigma}_{0}+\omega\dot{\alpha}^{2}\right].
\end{equation}

Now to kill the cross-term $\dot{\alpha}\dot{\sigma}_{0}$, we do a change of variables as, (which is in fact a canonical one) \footnote{\textit{We thank Ott Vilson for pointing out to us that this is a canonical transformation even at quantum level.}}
\begin{eqnarray}
\label{conformal}\beta_{0}&=&\sigma_{0}+\frac{\alpha}{2},\\
\beta_{\pm}&=&\sigma_{\pm}.
\end{eqnarray}
which yield following Lagrangian
\begin{equation}
L_{J} = \frac{e^{-\frac{\alpha}{2}+ 3\beta_{0}}}{n}\left[-6\dot{\beta}_{0}^{2}+6\dot{\beta}_{+}^{2}+6\dot{\beta}_{-}^{2}+\frac{2\omega +3}{2}\dot{\alpha}^2\right],
\end{equation}
 and the corresponding Hamiltonian is given by 
\begin{equation}
H_{J} =  \frac{ne^{\frac{\alpha}{2} - 3 \beta_{0}}}{4}\left[-\frac{1}{6}(p_0^2-p_+^2-p_-^2)+\frac{2}{2\omega+3}p_\alpha^2\right],
\end{equation}
where $p_{0}$ and $p_{\pm}$ are momenta conjugate to $\beta_{0}$ and $\beta_{\pm}$ respectively while $p_{\alpha}$ is momentum conjugate to $\alpha$. \\

The transformation (\ref{conformal}) looks like the one which takes us from the Jordan to the Einstein frame. But at this stage, these are to be thought merely as a canonical transformation. Hence, the Lagrangian and Hamiltonian above describes Brans-Dicke theory in a canonically equivalent frame of the Jordan frame.\\

Upon variation with respect to $n$, we obtain Hamiltonian constraint, 
\begin{equation}\label{Ham}
\mathcal{H}=\left[-\frac{1}{6}(p_0^2-p_+^2-p_-^2)+\frac{2}{2\omega+3}p_\alpha^2\right]=0.
\end{equation}

If we now consider a canonical transformation $(\alpha,p_\alpha)$ to $(T,p_T)$ as
\begin{eqnarray}
T&=&\frac{\alpha}{p_\alpha}\ ,\\
p_T&=&\frac{p_{\alpha}^{2}}{2}\ ,
\end{eqnarray}
 then Eq.~(\ref{Ham}) can be rewritten as 
\begin{equation}
\mathcal{H}=-\frac{1}{6}\left(p_0^2-p_+^2-p_-^2-\frac{24}{2\omega+3}p_T\right)\ .
\end{equation}

In the absence of a properly oriented scalar time parameter in the theory, we have used the evolution of the scalar field as the relevant time parameter. The method is the same as that suggested by Vakili\cite{vakili2012scalar}. That $T$ has the proper orientation can be ascertained from its dependence on the cosmic time in the right direction. For a summary of the method, we refer to \cite{spal1}. \\

Upon quantization, we obtain the Wheeler-DeWitt equation as follows,
\begin{equation}\label{master equation}
\left[\frac{\partial^2}{\partial \beta_0^2}-\frac{\partial^2}{\partial \beta_+^2}-\frac{\partial^2}{\partial \beta_-^2} - i\frac{24}{2\omega+3} \frac{\partial}{\partial T}\right]\psi=0\ .
\end{equation}

In the Einstein frame, the transformed metric components are ${\bar{g}}_{\mu\nu} = \phi g_{\mu\nu}$, and the action is given by
\begin{equation}
A_E=\int d^4x \sqrt{-\bar{g}}\left[\bar{R}+\frac{2\omega+3}{2}\partial_\mu \xi \partial^\nu \xi \right],
\end{equation}
where $\xi = \ln\phi$. 
Again, Bianchi I metric is given by, 
\begin{equation}
ds^2 = \bar{n}^2(t)dt^2 - \bar{a}^2(t)dx^2-\bar{b}^2(t)dy^2-\bar{c}^2(t)dz^2.
\end{equation}
The Lagrangian can be written as 
\begin{equation}
L_E=\frac{e^{3r_0}}{\bar{n}}\left[-6(\dot{r}_{0}^2-\dot{r}_{+}^{2}-\dot{r}_{-}^2)+\frac{2\omega+3}{2}\dot{\xi}^2\right],
\end{equation}
where
\begin{eqnarray}
\bar{a}(t)&=&e^{r_0 +r_+ +\sqrt{3}r_-}\ ,\\
\bar{b}(t)&=&e^{r_0 +r_+ -\sqrt{3}r_-}\ ,\\
\bar{c}(t)&=&e^{r_0 -2r_{+}}\ ,
\end{eqnarray}
and we obtain the following Hamiltonian 
\begin{equation}
H_E = \frac{\bar{n}e^{ - 3 r_0}}{4}\left[-\frac{1}{6}(\bar{p}_0^2-\bar{p}_+^2-\bar{p}_-^2)+\frac{2}{2\omega+3}p_\xi^2\right]\ .
\end{equation}
Now varying the action with respect to $\bar{n}$, one obtains the Hamiltonian constraint $\mathcal{H_E}=\frac{e^{  3 r_0}}{\bar{n}}H_E =0$. Again with similar canonical transformations as
\begin{eqnarray}
\bar{T}&=&\frac{\xi}{p_\xi}\ ,\\
p_{\bar{T}}&=&\frac{p_\xi^2}{2}\ ,
\end{eqnarray}
the WDW equation can be written as
\begin{equation}\label{ee}
\left[\frac{\partial^2}{\partial r_0^2}-\frac{\partial^2}{\partial r_+^2}-\frac{\partial^2}{\partial r_-^2} -i\frac{24}{2\omega+3} \frac{\partial}{\partial \bar{T}}\right]\bar{\psi}=0\ .
\end{equation}

One can easily show that $r_i=\beta_i$ and $T=\bar{T}$ from canonical transformations done in both the frame work. Hence it is quite obvious to see that the WDW equation in both the frame is exactly same. It is a trivial exercise to check that the wave packet, obtained solving Eq.~(\ref{ee}), will be the same as that formed out of the solution of the Eq.~ (\ref{master equation}), the WDW equation in the Jordan frame. To be mathematically precise, in Einstein frame, the wave function $\psi_E$ is a mapping from the configuration space $\tilde{\mathcal{C}}$, parameterized by $r_i$ to the real line, in Jordan frame we have wave function $\psi_J$, a mapping from the configuration space $\mathcal{C}$, parameterized by $\beta_i$ to the real line. The equality $r_i=\beta_i$ implies  that $\mathcal{C}=\tilde{\mathcal{C}}$ and the same form of WDW equation tells us that the functional form of $\psi_E$ is same with that of $\psi_J$; in short, wave functions $\psi_E:\tilde{\mathcal{C}}\mapsto\mathbf{R}$ and $\psi_J:\mathcal{C}\mapsto\mathbf{R}$ are precisely same functions, living in the same Hilbert space. 

It further implies that not only the two formulations are equivalent in the formal structure of the WDW equation, but also the probabilities and expectation values of observables will be same. The expectation value for operator A can be found as 
$\int dr_i \psi_J^{*}A \psi_J  = \langle A_J\rangle$ and $\int d\beta_i \psi_E^{*}A\psi_E   = \langle A_E\rangle$, where A should be expressed as a function of $r_i$ or $\beta_i$ respectively, but they are same anyway. So the equivalence is not merely in the classical limit, but rather in a real quantum picture. It also deserves mention that the quantities $r_i$ and $\beta_i$ are unitarily related, one can get one set from the other an identity transformation. Moreover, Should one treat $\sigma_i$ as fundamental variable in Jordan frame, the unitary equivalence is preserved as $\sigma_i$ is also related to $\beta_i$ by a unitary transformation, to be precise $\exp\left(\imath\frac{\alpha}{2}P_{\sigma_{0}}\right)\sigma_{0}\exp\left(-\imath\frac{\alpha}{2} P_{\sigma_{0}}\right)=\beta_0$ and $\beta_{\pm}$ is related to $\sigma_{\pm}$ by identity transformation where $P_{\sigma_0}$ is the momentum conjugate to $\sigma_0$. 

We further make a remark that if in Eq.~(\ref{b1metric}) we put $a=b=c$, then in Eq.~(\ref{b1trans}), both $\sigma_+$ and $\sigma_-$ become zero and we recover the same result for a spatially flat isotropic cosmological model in chapter \ref{ch5:1}. 
\section{Generalizations of isotropic models with constant but nonzero spatial curvature} \label{sec:2}
Bianchi V and Bianchi IX models have constant spatial curvature, with a negative and a positive signature respectively and are thus anisotropic generalizations of open and closed isotropic models.

\subsection{Bianchi V}

The Bianchi-V metric is given by
\begin{equation}
ds^2 = n^2(t)dt^2 - a^2(t)dx^2-e^{2 m x}[b^2(t)dy^2+c^2(t)dz^2],
\end{equation}
where $m$ is a constant. We parametrize the scale factors in following manner,
\begin{eqnarray}
a(t)&=&e^{\sigma_0 },\\
b(t)&=&e^{\sigma_0 +\sqrt{3}(\beta_+ -\beta_-)},\\
c(t)&=&e^{\sigma_0 -\sqrt{3}(\beta_+ -\beta_-)},\\
\phi(t)& =& e^{\alpha}
\end{eqnarray}
 and write down the Lagrangian, which will have some cross-term like $\dot{\alpha}\dot{\sigma_{0}}$. Now to kill the cross-term, we define $\beta_{0}=\sigma_{0}+\frac{\alpha}{2}$, which is again a canonical transformation. This immediately recasts the Lagrangian in following form, 

\begin{equation}
L_{J} = \frac{e^{-\frac{\alpha}{2} + 3 \beta_0}}{n}\left[-6\dot{\beta}_{0}^2+6(\dot{\beta}_{+}-\dot{\beta}_{-})^2-6e^{-2\beta_0}n^2m^2+\frac{2\omega +3}{2}\dot{\alpha}^2 \right].
\end{equation}

The non-trivial part of this parametrization is that the canonical transformation required to make the Lagrangian in a diagonal form is essentially same as conformal transformation to the Einstein frame. There is no a priori reason for them to be the same. We emphasize that the conformal transformation is indeed canonical even, at quantum level.  \\

The corresponding Hamiltonian in the Jordan frame can be written as 
\begin{equation}
H_{J} =  \frac{ne^{\alpha / 2 - 3 \beta_0}}{24}\left[-(p_0^2-p_+^2-144m^2e^{4\beta_0})+\frac{12}{2\omega+3}p_\alpha^2\right].
\end{equation}
A variation of the action with respect to $n$ yields $\mathcal{H}=\frac{e^{-\alpha / 2 + 3 \beta}H}{n}=0$.

Now we consider a canonical transformation $(\alpha,p_\alpha)$ to $(T,p_T)$, as before, given by following,
\begin{eqnarray}
T&=&\frac{\alpha}{p_\alpha}\ ,\\
p_T&=&\frac{p_\alpha^2}{2}\ ,
\end{eqnarray}
 so that Wheeler de Witt equation becomes
\begin{equation}
\left[\frac{\partial^2}{\partial \beta_0^2}-\frac{\partial^2}{\partial \beta_+^2}-144m^2e^{4\beta_0} -i\frac{24}{2\omega+3} \frac{\partial}{\partial T}\right]\psi=0\ .
\end{equation}

In the Einstein frame the action is given by
\begin{equation}
A_E=\int d^4x \sqrt{-\bar{g}}\left[\bar{R}+\frac{2\omega+3}{2}\partial_\mu \xi \partial^\nu \xi \right]\ ,
\end{equation}
where $\xi = \ln\phi$. The  Bianchi V metric is given by
\begin{equation}
ds^2 = \bar{n}^2(t)dt^2 - \bar{a}^2(t)dx^2-e^{2mx}[\bar{b}^2(t)dy^2+\bar{c}^2(t)dz^2]\ .
\end{equation}
Now the Lagrangian can be written as 
\begin{equation}
L_E=\frac{e^{3r_0}}{\bar{n}}\left[-6\dot{r}_{0}^2+6(\dot{r}_{+}-\dot{r}_{-})^2-6e^{-2r_0}n^2m^2+\frac{2\omega+3}{2}\dot{\xi}^2\right]\ ,
\end{equation}
where
\begin{eqnarray}
\bar{a}(t)&=&e^{r_0 } ,\\
\bar{b}(t)&=&e^{r_0 +\sqrt{3}(r_+ - r_-)}\ ,\\
\bar{c}(t)&=&e^{r_0 -\sqrt{3}(r_+ - r_-)}\ ,
\end{eqnarray}
and the corresponding Hamiltonian is now of the form,
\begin{equation}
H_E = \frac{\bar{n}e^{ - 3 r_0}}{24}\left[-(\bar{p}_0^2-\bar{p}_+^2-144m^2e^{4r_0})+\frac{12}{2\omega+3}p_\xi^2\right]\ .
\end{equation}
A variation of the action with respect to $n$ yields $\mathcal{H}_{E}=\frac{e^{  3 r_0}}{\bar{n}}H_E =0$. Again with similar canonical transformation, 
\begin{eqnarray}
\bar{T}=\frac{\xi}{p_\xi}\ ,\\
p_{\bar{T}}=\frac{p_\xi^2}{2}\ ,
\end{eqnarray}
Wheeler-DeWitt equation can be recast in following form, 
\begin{equation}
\left[\frac{\partial^2}{\partial r_0^2}-\frac{\partial^2}{\partial r_+^2}-144m^2e^{4r_0} -i\frac{24}{2\omega+3} \frac{\partial}{\partial \bar{T}}\right]\bar{\psi}=0\ .
\end{equation}

It is evident from the canonical transformation done in both the framework that $r_i=\beta_i$ and $T=\bar{T}$, hence the WDW equation in both the frame comes out to be exactly same and leads to similar wave packet if we transform  $\bar{a},\bar{b},\bar{c}$ back to the original $\sqrt{\phi} a,\sqrt{\phi} b,\sqrt{\phi} c$ respectively.

\subsection{Bianchi IX}
A Bianchi-IX metric given as
\begin{equation}
ds^2 = n^2(t)dt^2 - a^2(t)dr^2-b^2(t)d \theta^2 -[a^2(t) \cos^2 \theta+b^2(t)\sin^2 \theta]d \phi^2\ .
\end{equation}

In the Einstein frame, the metric components are indicated with an overhead bar. \\

With the following change of variables
 \begin{eqnarray}
b(t)&=&e^{-\alpha} \frac{\beta}{a}\ ,\\
a(t)&=&e^{-\alpha/2 } a_0\ ,\\
\phi(t) &=& e^{\alpha}\ ,
\end{eqnarray}
in the Jordan frame and following change 
\begin{equation}
 \bar{\beta}=\bar{a}\bar{b}\ ,
\end{equation}
in the Einstein frame coupled with a canonical transformation as done in the previous two cases, one can write down the WDW equations in the two frames. The result is the same, i.e., similar wave packet will be obtained if we revert the $\bar{a},\bar{b}$ to original $\sqrt{\phi} a, \sqrt{\phi} b$ respectively.

\section{Models with varying spatial curvature with no isotropic analogue} \label{sec:3}

The equivalence of the two frames can also be shown in other Bianchi models as well. Bianchi I, V and IX are in fact examples of anisotropic models which reduce to isotropic Friedmann models under given condition. But the conclusion remains the same for other models which do not have this property. 

\subsection{Bianchi VI model}
The example of a Bianchi type VI model can be taken up. Such a model is given by the metric 

\begin{equation}
ds^2 = n^2(t)dt^2 - a^2(t)dx^2-e^{- m x}b^2(t)dy^2 -e^{x}c^2(t)dz^2.
\end{equation}
With the change of variables as 
\begin{eqnarray}
a(t)&=&e^{-\alpha/2 +\beta_0 }\ ,\\
b(t)&=&e^{-\alpha/2 +\beta_0 +\sqrt{3}(\beta_+ -\beta_-)}\ ,\\
c(t)&=&e^{-\alpha/2 +\beta_0 -\sqrt{3}(\beta_+ -\beta_-)}\ ,\\
\phi(t)&=& e^{\alpha}\ ,
\end{eqnarray}

the Lagrangian and the Hamiltonian in the Jordan frame can be written respectively as
\begin{equation}
L_J = \frac{e^{-\alpha / 2 + 3 \beta_0}}{n}\left[-6\dot{\beta}_0^2+6(\dot{\beta}_+  \\
-\dot{\beta}_-)^2-\frac{e^{-2\beta_0}n^2(m^2-m+1)}{2}+\frac{2\omega +3}{2}\dot{\alpha}^2 \right]\ ,
\end{equation}
and
\begin{equation}
H_J =  \frac{ne^{\alpha / 2 - 3 \beta_0}}{24}\left[-[p_0^2-p_+^2-12(m^2-m+1)e^{4\beta_0}]+\frac{12}{2\omega+3}p_\alpha^2\right]\ .
\end{equation}

If we consider a canonical transformation $(\alpha,p_\alpha)$ to $(T,p_T)$ as
\begin{eqnarray}
T&=&\frac{\alpha}{p_\alpha}\ ,\\
p_T&=&\frac{p_\alpha^2}{2}\ ,
\end{eqnarray}
then with the Hamiltonian constraint, $\mathcal{H}=\frac{e^{-\alpha / 2 + 3 \beta}H}{n}=0$, the WDW equation can be written as
\begin{equation}
\label{b6WDW-j}
\left[\frac{\partial^2}{\partial \beta_0^2}-\frac{\partial^2}{\partial \beta_+^2}-12(m^2-m+1)e^{4\beta_0} -i\frac{24}{2\omega+3} \frac{\partial}{\partial T}\right]\psi=0\ .
\end{equation}
In the Einstein frame, the metric is written as
\begin{equation}
ds^2 = \bar{n}^2(t)dt^2 - \bar{a}^2(t)dx^2-e^{-mx}\bar{b}^2(t)dy^2-e^{x}\bar{c}^2(t)dz^2\ ,
\end{equation}
where the barred metric components are related to the unbarred components in the Jordan frame as ${\bar g}_{\mu\nu} = \phi g_{\mu\nu}$. With the transformation
\begin{eqnarray}
\bar{a}(t)&=&e^{r_0 }\ ,\\
\bar{b}(t)&=&e^{r_0 +\sqrt{3}(r_+ -r_-)}\ ,\\
\bar{c}(t)&=&e^{r_0 -\sqrt{3}(r_+ -r_-)}\ ,
\end{eqnarray}

The Lagrangian and the Hamiltonian in the Einstein frame look respectively as
\begin{equation}
L_E=\frac{e^{3r_0}}{\bar{n}}\left[-6\dot{r}_0^2+6(\dot{r}_+-\dot{r}_-)^2-\frac{e^{-2r_0}n^2(m^2-m+1)}{2}+\frac{2\omega+3}{2}\dot{\xi}^2\right]\ ,
\end{equation}
and 
\begin{equation}
H_E = \frac{\bar{n}e^{ - 3 r_0}}{24}\left[-[\bar{p}_0^2-\bar{p}_+^2-12(m^2-m+1)e^{4r_0}]+\frac{12}{2\omega+3}p_\xi^2\right]\ .
\end{equation}

Again with similar canonical transformation as
\begin{eqnarray}
\bar{T}&=&\frac{\xi}{p_\xi}\ ,\\
p_{\bar{T}}&=&\frac{p_\xi^2}{2}\ ,
\end{eqnarray}
and the Hamiltonian constraint  $\mathcal{H_E}=\frac{e^{  3 r_0}}{\bar{n}}H_E =0$, the WDW equation can be written as
\begin{equation}
\left[\frac{\partial^2}{\partial r_0^2}-\frac{\partial^2}{\partial r_+^2}-12(m^2-m+1)e^{4r_0} -i\frac{24}{2\omega+3} \frac{\partial}{\partial \bar{T}}\right]\bar{\psi}=0\ .
\end{equation}
One can easily show that $r_i=\beta_i$ and $T=\bar{T}$ from canonical transformation effected in both the framework, thus the WDW equation in both the frame is exactly same, henceforth leads to similar the wave packet upon transforming $\bar{a},\bar{b},\bar{c}$ back to original $a, b, c$ respectively as usual.

\subsection{Bianchi-I with a local rotational symmetry}
A locally rotationally symmetric (LRS) Bianchi -I model can be written in terms of the metric
\begin{equation}
\label{lrsI-metric}
ds^2 = n^2dt^2 - a^2(t)dx^2-b^2(t)e^{2x/h}(dy^2+dz^2)\ ,
\end{equation}
where $n(t)$ is the lapse function, $a(t), b(t)$ are functions of time and $h$ is a constant. The Ricci scalar can be written for this metric as 
\begin{multline}
 \sqrt{-g}\ \phi R = -e^{\frac{2x}{h}}\Bigg[\frac{6b^2\phi n}{h^2 a}-\frac{4b\phi \dot{a}\dot{b}}{n}-\frac{2a\phi\dot{b}^2}{n}+\frac{2b^2\phi\dot{a}\dot{n}}{n^2} +\frac{4ab\phi\dot{b}\dot{n}}{n^2}\\-\frac{2b^2\phi\ddot{a}}{n} -\frac{4ab\dot{b}\dot{\phi}}{n}\Bigg].
\end{multline}
Lagrangian in the Jordan frame can be written as 
\begin{equation}
L_J=\frac{b^2\phi a}{n}\left[-\frac{6n^2}{h^2a^2}-\frac{4\dot{a}\dot{b}}{ab}-\frac{2\dot{b}^2}{b^2}-\frac{2\dot{a}\dot{\phi}}{a\phi}-\frac{4\dot{b}\dot{\phi}}{b\phi}+\omega\frac{\dot{\phi}^2}{\phi^2} \right].
\end{equation}
With change of variable given as 
\begin{eqnarray}
a(t)&=&e^{-\alpha/2 -\beta_+ }\ ,\\
b(t)&=&e^{-\alpha/2 +\beta_+ +\beta_-}\ ,\\
\phi(t)&=& e^{\alpha}\ ,
\end{eqnarray}
the Lagrangian assumes the form 
\begin{equation}
L_J=\frac{e^{\beta_+ +2\beta_--\alpha/2}}{n}\left[2\dot{\beta}_{+}^2-2\dot{\beta}_{-}^2+\frac{2\omega+3}{2}\dot{\alpha}^2\right]-\frac{6 e^{\frac{\alpha}{2}+3 \beta_+ +2 \beta_-} n}{h^2}\ ,
\end{equation}
and corresponding Hamiltonian becomes
\begin{equation}
H_J =  \frac{ne^{\alpha/ 2 -\beta_+-2\beta_-}}{8}\left[p_{\beta_+}^2-p_{\beta_-}^2+\frac{4}{2\omega+3}p_\alpha^2+\frac{48e^{4(\beta_+ +\beta_-)}}{h^2}\right].
\end{equation}
On varying of the action with respect to $n$, we get  the Hamiltonian constraint $\mathcal{H}=\frac{e^{\beta_+ +2\beta_--\alpha/2}H}{n}=0$.

Now we consider a canonical transformation $(\alpha,p_\alpha)$ to $(T,p_T)$, as before, given by following relations,
\begin{eqnarray}
T&=\frac{\alpha}{p_\alpha}\ ,\\
p_T&=\frac{p_\alpha^2}{2}\ .
\end{eqnarray}
 so that Wheeler de Witt equation $\mathcal{H}\psi= 0$ becomes
\begin{equation}
\left[-\frac{\partial^2}{\partial\beta_+^2}+\frac{\partial^2}{\partial\beta_-^2}+\frac{48e^{4(\beta_+ +\beta_-)}}{h^2}-i\frac{8}{2\omega+3}\frac{\partial}{\partial T}\right]\psi=0.
\end{equation}

If we now transform the metric to the Einstein frame via a conformal transformation ${\bar g}_{\mu\nu} = \phi g_{\mu\nu}$, the metric looks like 
\begin{equation}
ds^2 = \bar{n}^2dt^2 - \bar{a}^2(t)dx^2-\bar{b}^2(t)e^{2x/h}(dy^2+dz^2)\ ,
\end{equation}
we can write the action in the Einstein frame as 
\begin{equation}
A_E=\int d^4x \sqrt{-\bar{g}}\left[\bar{R}+\frac{2\omega+3}{2}\partial_\mu \xi \partial^\nu \xi \right],
\end{equation}
where $\xi = ln\left(\phi\right)$ and the barred quantities indicate that they are in the Einstein frame. \\
With the transformation
\begin{eqnarray}
\bar{a}(t)&=&e^{-r_+ }\ ,\\
\bar{b}(t)&=&e^{r_+ +r_-}\ .
\end{eqnarray}
the Lagrangian and the Hamiltonian in the Einstein frame look respectively as
\begin{equation}
L_E=\frac{e^{r_++2r_-}}{n}\left[-2\dot{r}_{-}^2+2\dot{r}_{+}^2+\frac{2\omega+3}{2}\dot{\xi}^2\right]-\frac{6 e^{2 r_- +3 r_+} \bar{n}}{h^2}\ ,
\end{equation}
\begin{equation}
H_E = \frac{\bar{n}e^{-r_+ -2r_-}}{8}\left[\bar{p}_{r_+}^2-\bar{p}_{r_-}^2+\frac{4}{2\omega+3}p_\xi^2 +\frac{48e^{4(r_+ +r_-)}}{h^2}\right]\ .
\end{equation}
A variation of the action with respect to $\bar{n}$ yields $\mathcal{H}=\frac{e^{r_+ +2r_-}H}{\bar{n}}=0$.\\
Now we consider a canonical transformation of variables $(\xi,p_\xi)$ to $(\bar{T},p_{T})$, as before, given by following,
\begin{eqnarray}
T&=&\frac{\xi}{p_\xi},\\
p_T&=&\frac{p_\xi^2}{2},
\end{eqnarray}
 so that Wheeler-DeWitt equation $\mathcal{H}\bar{\psi}= 0$ becomes
\begin{equation}
\left[-\frac{\partial^2}{\partial r_+^2}+\frac{\partial^2}{\partial r_-^2}+\frac{48e^{4(r_+ +r_-)}}{h^2}-i\frac{8}{2\omega+3}\frac{\partial}{\partial \bar{T}}\right]\bar{\psi}=0.
\end{equation}
It is evident from the canonical transformation done in both the framework that $r_+=\beta_+,r_-=\beta_-$ and $T=\bar{T}$, hence the WDW equations in the two frames comes out to be exactly same and leads to similar wave packet if we transform  $\bar{a},\bar{b}$ back to the original $\sqrt{\phi} a,\sqrt{\phi} b$, respectively.

\subsection{Kantowski-Sachs model}

The Kantowski-Sachs (KS) metric is written as
\begin{equation}
ds^2 = n^2dt^2 - a^2(t)dx^2-b^2(t)(d\theta^2+sin^2 \theta d\Phi^2).
\end{equation}

The Lagrangian in the Jordan frame is
\begin{equation}
L_J=\frac{b^2\phi a}{n}\left[\frac{2n^2}{b^2}-\frac{4\dot{a}\dot{b}}{ab}-\frac{2\dot{b}^2}{b^2}-\frac{2\dot{a}\dot{\phi}}{a\phi}-\frac{4\dot{b}\dot{\phi}}{b\phi}+\omega\frac{\dot{\phi}^2}{\phi^2} \right].
\end{equation}

With a change of the set of variables given as 
\begin{eqnarray}
a(t)&=&e^{-\alpha/2 -\beta_+ },\\
b(t)&=&e^{-\alpha/2 +\beta_+ +\beta_-},\\
\phi(t)&=& e^{\alpha},
\end{eqnarray}
the Lagrangian can be re-written as 
\begin{equation}
L_J=\frac{e^{\beta_+ +2\beta_--\alpha/2}}{n}\left[2\dot{\beta}_+^2-2\dot{\beta}_-^2+\frac{2\omega+3}{2}\dot{\alpha}^2\right]+2n  e^{\frac{\alpha}{2}- \beta_+ } 
\end{equation}
and the corresponding Hamiltonian looks like
\begin{equation}
H_J =  \frac{ne^{\alpha/ 2 -\beta_+-2\beta_-}}{8}\left[p_{\beta_+}^2-p_{\beta_-}^2+\frac{4}{2\omega+3}p_\alpha^2-16e^{2\beta_-}\right].
\end{equation}

A variation of the action with respect to $n$ yields $\mathcal{H}=\frac{e^{\beta_+ +2\beta_--\alpha/2}H}{n}=0$.\\
Now we consider a canonical transformation $(\alpha,p_\alpha)$ to $(T,p_T)$, as before, given by following,
\begin{eqnarray}
T&=&\frac{\alpha}{p_\alpha}\ ,\\
p_T&=&\frac{p_\alpha^2}{2}\ ,
\end{eqnarray}
 so that Wheeler-DeWitt equation $\mathcal{H}\psi= 0$ becomes
\begin{equation}
\left[-\frac{\partial^2}{\partial\beta_+^2}+\frac{\partial^2}{\partial\beta_-^2}-16e^{2\beta_-}-i\frac{8}{2\omega+3}\frac{\partial}{\partial T}\right]\psi=0\ .
\end{equation}

In the Einstein frame, the transformed metric is  
\begin{equation}
ds^2 = \bar{n}^2dt^2 - \bar{a}^2(t)dx^2-\bar{b}^2(t)(d\theta^2+sin^2 \theta d\Phi^2)\ ,
\end{equation}
where the barred metric components are related to the unbarred components in the Jordan frame as ${\bar g}_{\mu\nu} = \phi g_{\mu\nu}$.

With the transformations
\begin{eqnarray}
\bar{a}(t)&=&e^{-r_+ },\\
\bar{b}(t)&=&e^{r_+ +r_-},
\end{eqnarray}

the Lagrangian and the Hamiltonian in the Einstein frame look respectively as
\begin{equation}
L_E=\frac{e^{r_++2r_-}}{n}\left[2\dot{r}_{+}^2-2\dot{r}_{-}^2+\frac{2\omega+3}{2}\dot{\xi}^2\right]+2\bar{n}e^{-r_+} ,
\end{equation}
\begin{equation}
H_E = \frac{\bar{n}e^{-r_+ -2r_-}}{8}\left[\bar{p}_{r_+}^2-\bar{p}_{r_-}^2+\frac{4}{2\omega+3}p_\xi^2 -16e^{2r_-}\right].
\end{equation}
A variation of the action with respect to $\bar{n}$ yields $\mathcal{H}=\frac{e^{r_+ +2r_-}}{\bar{n}}H=0$.\\
Now we consider a canonical transformation $(\xi,p_\xi)$ to $(\bar{T},p_{T})$, as before, given by following,
\begin{eqnarray}
\bar{T}&=&\frac{\xi}{p_\xi},\\
p_{\bar{T}}&=&\frac{p_\xi^2}{2},
\end{eqnarray}
so that Wheeler-DeWitt equation $\mathcal{H}\bar{\psi}= 0$ becomes
\begin{equation}
\left[-\frac{\partial^2}{\partial r_+^2}+\frac{\partial^2}{\partial r_-^2}-16e^{2r_-}-i\frac{8}{2\omega+3}\frac{\partial}{\partial \bar{T}}\right]\bar{\psi}=0.
\end{equation}
It is evident from the canonical transformation done in both the framework that $r_+=\beta_+,r_-=\beta_-$ and $T=\bar{T}$, hence the WDW equation in both the frame comes out to be exactly same and leads to similar wave packet if we transform  $\bar{a},\bar{b}$ back to the original $\sqrt{\phi} a,\sqrt{\phi} b$ respectively.

\section{Generic Scenario and Concluding Remarks}

By a mathematical equivalence, we mean any physical quantity obtained in the Jordan frame can be mathematically transformed into an expression in the Einstein frame such that the expression is exactly the same in both the frames.\\

Quantum mechanically, this mathematical equivalence seems to be broken as reported in the literature several times\cite{cho1992, nb}. Here we show that the reported nonequivalence can be restored with a consistent operator ordering. Given a frame, be it the Einstein or the Jordan, we always have to choose a particular ordering of operators so as to go over to quantum theory. Now, choosing one particular operator ordering in the Jordan frame fixes the operator ordering in the Einstein frame and vice versa. This strips us off the freedom of choosing operator ordering in both the frames independently. Should we do the ordering in both the frames arbitrarily, it might become inconsistent with each other, leading to a different quantum Hamiltonian and hence different behavior of the wave packets. Thus, the apparent discrepancy should not be attributed to quantum effects, rather they are the artifacts of an inconsistent operator ordering. In this work, we  basically have picked up a consistent choice of operator ordering in both the frames. We do the parametrization in a suitable way, which renders the most natural choice of operator ordering to be the consistent ones\footnote{To avoid any potential confusion, the operator ordering we talk about is not the normal ordering in quantum field theory, and this also does not have anything to do with the field redifinition or frame transformation. As we mentioned in the introduction, the ordering issue comes as we promote the classical Himaltonian to a quantum one, position and momenta become non-commuting operators. The important thing is that we choose an ordering in one frame, and the ordering should be consistent with that in the other frame. The initial choice could be one in which the theory preserves unitarity\cite{spaljmp}.}. It worths mentioning though that, \cite{nb} deals with a cosmological model with a matter content. Unless the matter content is a conformal one, i.e, radiation, the equivalence is broken explicitly. It would be interesting to show that with a consistent operator ordering at least for a conformal fluid, i.e., radiation, equivalence can be restored. \\

It deserves mention that the parametrization in this particular way and the natural operator ordering in the new variables $\beta_{0},\beta_{\pm}$ indicate a particular choice of operator ordering in terms of old variables $a,b,c$. This choice is by no means a unique one and it is not required to prove the uniqueness either. For homogeneous models, however, there is a natural unique  choice of the parametrization as shown in what follows.

Here we explicitly show that for every choice of co-ordinates and corresponding quantization in the Jordan frame with a particular operator ordering, we have a particular choice of co-ordinates and an operator ordering in the Einstein frame so that both the frames become equivalent. In the Jordan frame, we parametrize the metric in following way: 
\begin{equation}
g_{ij}= e^{-\alpha} h_{ij}, \quad\phi=e^{\alpha}.
\end{equation}

As we are dealing with homogeneous models, we parametrize $h_{ij}$ by some function of time in some particular way. Thus we have 
\begin{equation}
\mathcal{L}_{J}=ne^{-\frac{\alpha}{2}} \sqrt{h} \left[R_{h_{ij}} + \frac{2\omega +3}{2}\dot{\alpha}^{2}\right].
\end{equation}

From this, we write down the Hamiltonian. Subsequently the Hamiltonian constraint is obtained and the theory is quantized in a canonical way.\\

Now, the Einstein frame is a conformal transformation of the Jordan frame. The homogeneity guarantees that $\phi$ is a function of $t$ alone, hence the conformal factor is a function of $t$ alone. This implies that going from the Jordan to the Einstein frame does not change the Bianchi class of the model considered. Had it been the case that $\phi$ is a function of space as well as time, things would have been more intricate and this method would have broken down, i.e., the proof would not have gone through.\\

 In the Einstein frame, we parametrize the metric $\bar{g}_{ij}$ in same way as we have parametrized the metric $h_{ij}$. This will lead to
\begin{equation}
\mathcal{L}_{E}=\bar{n}\sqrt{-\bar{g}}\left[\bar{R}+\frac{2\omega +3}{2}\dot{\alpha}^{2}\right]
\end{equation}

Since, $\bar{g}_{ij}=\phi g_{ij}=h_{ij}$, we have $\bar{R}=R_{h_{ij}}$, and the parametrization in a similar way guarantees that a consistent operator ordering is being chosen automatically, hence the equivalence becomes an obvious one. This is to be emphasized that we have shown equivalence for every choice of parametrization of $h_{ij}$ in this work. \\

It deserves mention that this proof is really independent of how we are defining the time variable for quantum theory, only a properly oriented time parameter is required.  Nonetheless, one might wonder that the choice of parametrization or the transformation, given by Eq.~(\ref{conformal}) to make the Lagrangian diagonal is actually effecting a conformal transformation to Einstein frame in disguise. However, this transformation can be thought of as a canonical one. Since, a canonical transformation preserves the quantum structure, the two frames are indeed equivalent. \\

It is a straight forward exercise to extend the prescription for a cosmological model with fluid. It deserves mention that the application of \textit{ S-theorem} \cite{spal-scale} and resulting anomalous symmetry breaking in quantum FRW cosmological model with radiation matter content \cite{spal5} is technically done using the same transformation. Thus, it is worthwhile to check whether such symmetry breaking can happen in the Jordan frame as well before effecting the canonical transformation and confirm the quantum equivalence. \\

We point out that we have used the so called Misner variables\cite{misner1969quantum} in writing the metric. In these coordinates, a lot of simplification is achieved for anisotropic but spatially homogeneous spaces, namely the spacetimes that fall under the Bianchi classification. As general relativity is diffeomorphism invariant, this simplification does not lead to any change in physical properties. In fact, it appears that Misner coordinates are arguably the best set of coordinates for the purpose of quantization. We refer to the work of Agostini, Cianfrani and Montani for a very recent application of Misner variables\cite{agostini}, which indicates that Misner variables and other approaches do yield equivalent results for isotropic matter distribution. In the present case, although the geometry is anisotropic, the matter distribution is indeed isotropic.\\
 
It also deserves mention that we have tried other transformations involving the Brans-Dicke parameter $\omega$, and at least for some particular vaules of $\omega$ could write down separable expressions. One  difficulty is that of picking up a nice time variable, whose corresponding momentum appears only in the first order. But the most important problem is that it is impossible to pick the corresponding operator ordering in the two frames, which is irrecoverably lost in the transformation.\\

Indeed this equivalence has been already shown to be true in isotropic models as in the previous chapter. But isotropic models are special, the self-adjoint extension is quite unique in such models\cite{spaljmp}. But anisotropic models are not as simple, the uniqueness of  self-adjoint extension is not guaranteed, and the proper operator ordering is quite nontrivial.  So the present work strengthens the present knowledge regarding the equivalence between conformally transformed versions of the same theory. \\

We hope that the work will clarify the issue regarding equivalence prevailing in the literature. The question of mathematical equivalence, thus being clarified, one could ask for more sophisticated questions like quantum mechanically which frame is more useful and how the quantum behavior changes once we go from one frame to another frame. There is already some work in this connection\cite{cho1992}. 


\chapter{Conclusion}

Quantum cosmology makes an  attempt to apply the principles of quantum mechanics to cosmological systems. The quantum description of the universe is required for the early stage of the evolution at an energy scale where classical gravity loses its viability. The quantum prescription requires a suitable time parameter for the evolution of the system as cosmic time t is a coordinate itself in the general theory of relativity. Schutz's formalism  help us in this regard as discussed in Section \ref{ch:intro4}. It is also believed that anisotropic quantum cosmological models suffer from the problem of a non-unitary evolution. This alleged problem of non-unitarity has recently been resolved for a few anisotropic models with constant spatial curvature by showing that the WDW quantization scheme can lead to either the self-adjoint Hamiltonian or the Hamiltonian admitting self-adjoint extension by a suitable operator ordering \cite{spal1,spal2}. In chapter \ref{ch2:1}, we used the WDW quantization scheme to quantize the two examples of anisotropic cosmological models with varying spatial curvature, namely Bianchi II and VI. We showed the self-adjoint extension of Hamiltonian for these models is quite possible; thus, they do admit a unitary evolution. 
 
 In chapter \ref{ch3:1}, we tried to generalize the work of unitary evolution to higher dimensional anisotropic models. We quantize the n-dimensional anisotropic model with fluid. Using the WDW quantization method, we obtained the finite normed and time independent wave packet for stiff fluid $\alpha=1$. Thus, the unitary evolution of the model is established. We also computed the expectation values of scale factors and were able to show that they were non-singular. The expectation value of the volume element has non-zero minima, which indicates a bouncing universe. Thus, the model avoids the problem of singularity similar to earlier work done for anisotropic models in the absence of fluid \cite{ndaniso}. Despite the mathematical difficulty of the model with general fluid $\alpha \neq 1$, we were able to exhibit the unitary evolution  for n=7.

The unitary evolution of anisotropic models is achieved mainly through self-adjoint extension. In chapter \ref{ch4:1}, we tried to investigate the cost of doing such extension either by operator ordering or transformation of variables at the classical level. We considered the Bianchi-I cosmological model as an example. We showed that Noether symmetry is preserved in the process of self-adjoint extension. We also showed that scale invariance is lost while achieving the unitarity of the model. It has already been shown that the unitary evolution is not achieved at the cost of anisotropicity itself \cite{spal4}. 

Since unitary evolution has been extensively shown in this thesis for anisotropic but homogeneous models, the more difficult task would be to quantize the inhomogeneous cosmological models as it becomes challenging to do a 3+1 decomposition and apply the WDW quantization scheme in inhomogeneous models, even in the simpler model like Lemaitre-Tolman-Bondi (LTB) model. One may attempt to quantize such models in the future. 

In the latter part of the thesis, we tried to quantize the Brans-Dicke theory, which has been studied in two frames, the Jordan Frame and The Einstein frame. We tried to address the question of the equivalence of these two frames at the quantum level in chapter \ref{ch5:1}.  We quantized the spatially homogeneous and isotropic spacetime. We constructed the time parameter from the scalar field in the absence of the fluid in the model and obtained the wave packet in both the frames. Then we reverted the conformal transformation, $\bar{g}_{\mu\nu}=\Omega^2g_{\mu\nu}$ through which these two frames are related to each other. The wave packet obtained in the Einstein matched precisely with the one in the Jordan frame. Thus, we were able to show the equivalence of these two frames very clearly. 

After showing the equivalence of the Jordan and the Einstein frame for the isotropic quantum model, we tried to explore the equivalence of these two frames in anisotropic quantum cosmological models. In chapter \ref{ch6:1}, we first take the example of Bianchi I space time and derive the WDW equation in both the frames using various canonical transformation. The same form of the WDW equation in both the frames tells the functional form of the wave packet in the Einstein frame $\Psi_E$ is the same to one in the Jordan frame $\Psi_J$. We also remark that this equivalence is not only a mathematical construct, rather a real physical one, as the expectation value of an operator A would be same in these two frames.

Then we tried to extend this equivalence to various other anisotropic models, models with constant but non-zero spatial curvature namely Bianchi V and Bianchi IX model and models with varying spatial curvature, namely Bianchi VI, Bianchi I with local rotational symmetry and  Kantawski-Sachs model.

At the end of chapter \ref{ch6:1}, we show that we have a particular choice of coordinates transformations and an operator ordering in the Einstein frame for every such choice in the Jordan frame, which leads to the equivalence of the frames for any homogeneous model. We also emphasize on the fact that transformations are canonical, which preserves the quantum structure. Thus, two frames are equivalent at quantum level.

As a future work, one can look at this equivalence in the presence of matter, such as a fluid. The crucial canonical transformation related with these two frames, being conformal in nature, may preserve the equivalence for conformally invariant matter fields like trace-free fields such as a radiation fluid or an electromagnetic field.   
  
\backmatter
\begin{titlepage}
\newpage
\clearpage
\newpage
\mbox{~}
\clearpage
\newpage
\end{titlepage}
\bibliographystyle{myapsrev}
\addcontentsline{toc}{chapter}{Bibliography}
\bibliography{dissertation}

\end{document}